\documentclass[conference]{IEEEtran}
\IEEEoverridecommandlockouts
\usepackage{multirow}
\usepackage{subcaption}
\usepackage{cite}
\usepackage{diagbox}
\usepackage{amsmath,amssymb,amsfonts}
\usepackage{algorithmic}
\usepackage{graphicx}
\usepackage{textcomp}
\usepackage{xcolor}
\usepackage{booktabs}
\usepackage{url}

\makeatletter

\newcommand{\Rmnum}[1]{\expandafter\@slowromancap\romannumeral #1@}
\makeatother

\def\BibTeX{{\rm B\kern-.05em{\sc i\kern-.025em b}\kern-.08em
    T\kern-.1667em\lower.7ex\hbox{E}\kern-.125emX}}
\begin{document}

\title{Reinforced Path Reasoning for Counterfactual Explainable Recommendation\\
\thanks{This work is supported by the Australian Research Council (ARC) under Grant No. DP200101374, LP170100891, DP220103717 and LE220100078.}
}

\author{\IEEEauthorblockN{Xiangmeng Wang}
\IEEEauthorblockA{\textit{Data Science and Machine Intelligence Lab} \\
\textit{University of Technology Sydney}\\
Sydney, Australia \\
xiangmeng.wang@student.uts.edu.au}
\and
\IEEEauthorblockN{Qian Li \IEEEauthorrefmark{1}\thanks{\IEEEauthorrefmark{1}: Contributing equally with the first author}\IEEEauthorrefmark{2}}
\IEEEauthorblockA{\textit{School of Elec Eng, Comp and Math Sci} \\
\textit{Curtin University}\\
Perth, Australia \\
qli@curtin.edu.au}
\and
\IEEEauthorblockN{Dianer Yu}
\IEEEauthorblockA{\textit{Data Science and Machine Intelligence Lab} \\
\textit{University of Technology Sydney}\\
Sydney, Australia \\
Dianer.Yu-1@student.uts.edu.au}
\and
\IEEEauthorblockN{Guandong Xu\IEEEauthorrefmark{2}\thanks{\IEEEauthorrefmark{2}: Corresponding author}}
\IEEEauthorblockA{\textit{Data Science and Machine Intelligence Lab} \\
\textit{University of Technology Sydney}\\
Sydney, Australia \\
guandong.xu@uts.edu.au}
}

\maketitle

\begin{abstract}
Counterfactual explanations interpret the recommendation mechanism via exploring how minimal alterations on items or users affect the recommendation decisions. 
Existing counterfactual explainable approaches face huge search space and their explanations are either action-based  (e.g., user click) or aspect-based (i.e., item description).
We believe item attribute-based explanations are more intuitive and persuadable for users since they explain by fine-grained item demographic features (e.g., brand).
Moreover, counterfactual explanation could enhance recommendations by filtering out negative items.
  
In this work, we propose a novel \emph{Counterfactual Explainable Recommendation (CERec)} to generate item attribute-based counterfactual explanations meanwhile to boost recommendation performance. 
Our CERec optimizes an explanation policy upon uniformly searching candidate counterfactuals within a reinforcement learning environment.
We reduce the huge search space with an adaptive path sampler by using rich context information of a given knowledge graph.
We also deploy the explanation policy to a recommendation model to enhance the recommendation.
Extensive explainability and recommendation evaluations demonstrate CERec's ability to provide explanations consistent with user preferences and maintain improved recommendations.
We release our code at~\url{https://github.com/Chrystalii/CERec}.
\end{abstract}


\section{Introduction}

Modern recommendation models become sophisticated and opaque by modeling complex user/item context, e.g., social relations~\cite{li2022causal} and item profiles~\cite{hu2018leveraging}.
Hence the pressing need for faithful explanations to interpret user preferences meanwhile enable model transparency.
Explainable Recommendation Systems (XRS) aim to provide personalized recommendations complemented with explanations that answer why particular items are recommended~\cite{zhang2020explainable}.
It is fairly well-accepted that high-quality explanations can help improve users’ satisfactions and recommendation persuasiveness~\cite{wang2019explainable,xie2021explainable}.
Explanations can also facilitate system designers in tracking the decision-making of recommendation models for model debugging~\cite{DBLP:conf/aiia/FrosstH17}.

Existing XRS approaches can be categorized to model-intrinsic methods~\cite{shulman2020meta,lin2000collaborative} and model-agnostic methods~\cite{zhang2014explicit,ghazimatin2019fairy,wang2018reinforcement,peake2018explanation}.
Model-intrinsic methods seek recommendation models that are inherently interpretable, e.g., decision trees~\cite{shulman2020meta} and association rules~\cite{lin2000collaborative}, thus are limited in their applications to prominent deep learning models~\cite{zhang2020explainable}.
Recently, model-agnostic methods attract increasing attention~\cite{ai2021model} to allow the recommendation model to be black-box models, e.g., neural networks~\cite{ghazimatin2019fairy}.
Model-agnostic approaches identify correlations between input user-item interactions and output recommendations from black-box models~\cite{ai2021model}, to help users understand how their behaviors~\cite{ghazimatin2019fairy} (e.g., click) or which item features~\cite{wang2018reinforcement,DBLP:journals/corr/abs-1806-11330} (e.g., brand) contribute to the recommendation. 
However, existing model-agnostic approaches suffer from major limitations:
1) They should firstly identify influential factors (e.g., item features) that cause positive user interactions and then construct explanations, 
while they seldom consider what recommendations would be if we directly intervene on explanations to alternative ones.
2) They do not consider explanation complexity by constructing explanations with a fixed number of influential factors~\cite{zhang2014explicit}, while totally ignoring to search for a minimal set of influential factors that well-explain reasons for recommendations.

Recently, counterfactual explanation~\cite{tan2021counterfactual} has emerged as a favorable opportunity to solve the above questions.
Typically, \textit{counterfactual explanation} is defined as a minimal set of influential factors that, if applied, flip the recommendation decision.
Counterfactual explanation interprets the recommendation mechanism via exploring how minimal alterations on items or users affect the recommendation decisions.
To build counterfactual explanations, we have to fundamentally address:
\emph{what the recommendation result would be if a minimal set of factors (e.g., user behaviors/item features) had been different}~\cite{verma2020counterfactual}.
With counterfactual explanations, users can understand how minimal changes affect recommendations and conduct counterfactual thinking under ``what-if'' scenario~\cite{pearl2009causal}.

Existing counterfactual explanation based XRS can be summarized into two main categories.
The first line of search-based approaches~\cite{kaffes2021model,xiong2021counterfactual} conducts greedy search on counterfactual instances given by perturbing user interactions or item features.
Those counterfactual instances with the highest scores measured by heuristic quality metrics are considered counterfactual explanations. 
For example, Kaffes et al.~\cite{kaffes2021model} perturb counterfactual instances upon removal of items from users' interactions, with normalized length and candidate impotence measuring counterfactuals to guide the search. 
Xiong et al.~\cite{xiong2021counterfactual} perform constrained feature perturbations on item features and search perturbed features as counterfactual explanations. 
But search-based approaches suffer from high computational burden for the large-scale search space~\cite{verma2020counterfactual}.
Another line of optimization-based approaches~\cite{cheng2019incorporating,ghazimatin2020prince,tran2021counterfactual,tan2021counterfactual} formulates counterfactual reasoning as an optimization problem to alleviate the computation burden.
Typically, the optimization goal is to find simple but essential user actions or item aspects that directly cause user preference changes. 
Ghazimatin et al.~\cite{ghazimatin2020prince} perform random walks over a user-item interaction graph and calculate PageRank scores after removing user actions edges from the graph. 
User actions that change PageRank scores are considered as counterfactual explanations.
Tan et al.~\cite{tan2021counterfactual} modify item aspect scores to observe user preference changes based on a pre-defined user-aspect preference matrix.
In summary, existing optimization-based approaches either focus on user action~\cite{cheng2019incorporating,ghazimatin2020prince,tran2021counterfactual} or item aspect explanations~\cite{tan2021counterfactual},
leaving the item attribute-based counterfactual explanations largely unexplored.

\begin{figure}
    \centering
    \includegraphics[width=0.45\textwidth]{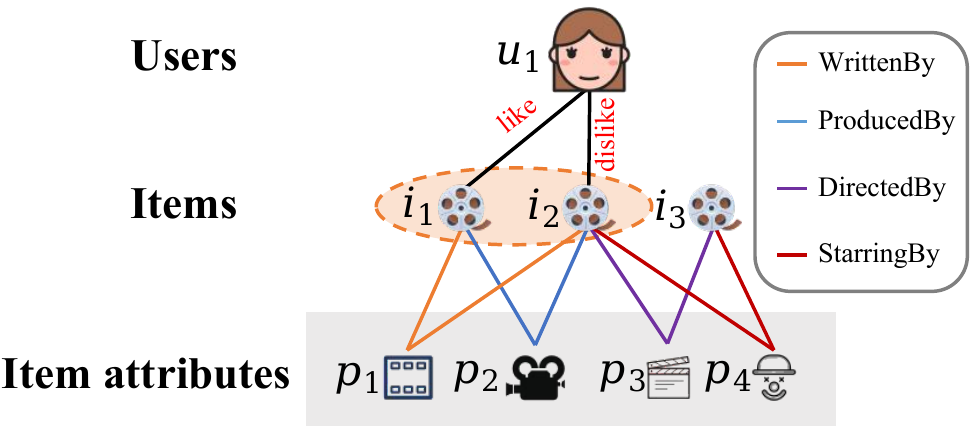}
    \caption{Toy example of inferring item attribute-based counterfactual explanations from knowledge graphs.}
    \label{fig:toy}
\end{figure}

We claim that item attribute-based counterfactual explanations could benefit both users' trust and recommendation performance.
On the one hand, item attribute-based counterfactual explanations are usually more intuitive and persuadable to seek users' trust.
This is mainly because users prefer to know detailed information, e.g., which item attribute caused the film ``Avatar'' not be recommended anymore? Is it the director of ``Avatar''?
Besides, it is essential to search for a minimal change on item attributes that would alter the recommendation. 
We take Figure~\ref{fig:toy} as an example.
$u_1$ gave negative feedback on $i_2$ with attributes $\{p_1,p_2,p_3,p_4\}$.
We could infer attributes $\{p_1,p_2,p_3,p_4\}$ reveal negative preferences of $u_1$.
However, item $i_1$ with attributes $\{p_1, p_2\}$ was liked by $u_1$, which somehow reflect positive user preferences on $\{p_1, p_2\}$.
Thus, merely using attributes $\{p_1,p_2,p_3,p_4\}$ as explanations for $u_1$'s recommendations would be controversial and misunderstand users' preferences.
In fact, the slightly changes between $i_1$'s attributes and $i_2$'s attributes, i.e., $\{p_3,p_4\}$, could be the true determinant factors for explaining $u_1$'s dislike.
On the other hand, 
counterfactual explanations could boost recommendation performance since they offer high-quality negative signals of user preferences.
Particularly, if we know the slightly changes $\{p_3,p_4\}$ explain $u_1$'s dislike, we could infer that item $i_3$ with $\{p_3, p_4\}$ would be disliked by $u_1$ as well.
To the end, the counterfactual explanation helps generate more precise recommendations by filtering out negative items.

In this work, we leverage knowledge graphs (KGs) that represent relations among real-world entities of users, items, and attributes to infer attribute-based counterfactual explanations meanwhile boosting recommendation performance. 
We propose a new \emph{Counterfactual Explainable Recommendation (CERec)} that crafts the counterfactual optimization problem as a reinforcement learning (RL) task.
The RL agent optimizes an explanation policy upon uniformly searching candidate counterfactuals.
To reduce the search space, we propose an adaptive path sampler over a KG with a two-step attention mechanism and select counterfactual item attributes as the explanation candidates. 
With explanation candidates, the RL agent optimizes the explanation policy to find optimal counterfactual attribute explanations.
We also deploy the explanation policy to a recommendation model to enhance the recommendation.
The contributions of our work are:

\begin{itemize}
    \item To the best of our knowledge, we are the first to leverage the rich attribute information in knowledge graphs to provide attribute-based counterfactual explanations for recommendations.
    
    \item We propose an RL-based framework to find the optimal counterfactual explanations, driven by an adaptive path sampler and a counterfactual reward function. 
    
    \item We design an adaptive path sampler with a two-step attention mechanism to reduce the search space of counterfactual explanations.
    
    \item We train the recommendation model uniformly with the counterfactual explanations with the aim of boosting the recommendation.
    
    \item Extensive results on explainability and recommendation performance demonstrate the effectiveness of CERec.
\end{itemize}

\section{Related work}

\subsection{Explainable Recommendation}

Explainable recommendation is proven to improve user satisfaction~\cite{tintarev2012evaluating} and system transparency~\cite{balog2020measuring}. 
Researches in explainable recommendation can be divided into two main categories~\cite{zhang2020explainable,gedikli2014should}. 
The first category of model-intrinsic approaches~\cite{shulman2020meta,lin2000collaborative,vig2009tagsplanations,herlocker2000explaining,sarwar2001item,papadimitriou2012generalized} designs interpretable recommendation models to facilitate explanations. These model-intrinsic methods train explainable models to identify influential factors for users' preferences, and then construct explanations accordingly. 
For example, Shulman et al.~\cite{shulman2020meta} propose a per-user decision tree model to predict users' ratings on items.
Each decision tree is built with a single user as a root and items as leaves, while a linear regression is applied to learn leaf node values to build decision rules.
The explanation is formed as a sequence of decisions along the decision tree.
Lin et al.~\cite{lin2000collaborative} propose a personalized association rule mining method targeted at a specific user.
Explanations are generated by identifying association rules between users and items.
Other works~\cite{vig2009tagsplanations,herlocker2000explaining,sarwar2001item,papadimitriou2012generalized,wang2022causal,10.1145/3533725} explore content-based approach~\cite{vig2009tagsplanations} and neighborhood-based collaborative filtering~\cite{herlocker2000explaining,sarwar2001item} to generate explanations based on users' neighbors~\cite{herlocker2000explaining,wang2022causal,10.1145/3533725}, item similarity~\cite{sarwar2001item}, user/item features~\cite{vig2009tagsplanations}, or combinations thereof~\cite{papadimitriou2012generalized}.

As modern recommendation models become complicated and black-box, e.g., embedding-based models~\cite{hamilton2017inductive,wang2019heterogeneous,hu2020heterogeneous} and deep neural networks~\cite{chen2018neural,he2017neural}, developing model-intrinsic approaches~\cite{zhang2020explainable} is not practical anymore.
Thus, another category of model-agnostic methods~\cite{ghazimatin2019fairy,DBLP:journals/corr/abs-1806-11330,nobrega2019towards} aims to explain black-box models in a post-hoc manner.
For instance, 
Ghazimatin et al.~\cite{ghazimatin2019fairy} perform graph learning on a heterogeneous information network that unifies users' social relations and historical interactions.
The learnt graph embeddings are used to train an explainable learning-to-rank model to output explanation paths.
Singh et al.~\cite{DBLP:journals/corr/abs-1806-11330} train a Learning-to-rank latent factor model to produce ranking labels. 
These ranking labels are then used for training an explainable tree-based model to generate interpretable item features for ranking lists.

\subsection{Counterfactual Explainable Recommendation}

Counterfactual explanations have a long history in philosophy, psychology, and social sciences ~\cite{verma2020counterfactual, DBLP:journals/corr/abs-2001-07417,DBLP:journals/corr/abs-1901-07694}.
They have been considered as satisfactory explanations ~\cite{woodwardMakingThingsHappen2004,duong2021stochastic,li2022deep} and elicit causal reasoning in humans~\cite{byrne2007rational,li2021causalOptimal,li2021causalAware}.
The counterfactual explanation has recently emerged as a hot topic in recommendation systems.
Successful works can be categorized into search-based and optimization-based approaches. 
The first category of search-based approaches~\cite{kaffes2021model,xiong2021counterfactual} performs greedy search for counterfactual explanations. 
For instance, Kaffes et al.~\cite{kaffes2021model} perturb user-item interactions by deleting items from user interaction queues to generate perturbed user interactions as counterfactual explanations.
Then a breadth-first search is used to find counterfactual explanations who achieve the highest normalized length and candidate impotence scores.
Xiong et al.~\cite{xiong2021counterfactual} propose a constrained feature perturbations on the features of candidate items and consider the perturbed item features as counterfactual explanations.
The second category of optimization-based approaches~\cite{cheng2019incorporating,ghazimatin2020prince,tran2021counterfactual,tan2021counterfactual} optimize explanation models to find counterfactual explanations with minimal changes. 
For instance, 
Ghazimatin et al.~\cite{ghazimatin2020prince} perform random walks over a heterogeneous information network and calculate the PageRank scores after removing user actions edges from the graph. 
Those minimal sets of user actions that change PageRank scores are deemed counterfactual explanations.
Tran et al.~\cite{tran2021counterfactual} identify a minimal set of user actions that updates parameters of neural models.
Tan et al.~\cite{tan2021counterfactual} modify item aspect scores to observe user preference changes based on a user-aspect preference matrix.

Existing optimization-based approaches either focus on user action~\cite{cheng2019incorporating,ghazimatin2020prince,tran2021counterfactual} or item aspect explanations~\cite{tan2021counterfactual}.
In contrast, our counterfactual explanation is defined on the item attribute-side based on real-world item demographic features. 
Besides, we combine the search-based and optimization-based methods into an integrated mechanism by searching candidate counterfactuals to optimize an explanation policy within a reinforcement learning environment.

\section{Preliminary}
In this section, we first describe the collaborative knowledge graph that unifies user-item interactions and an external knowledge graph.
Then, we introduce the concept of attribute-based counterfactual explanation for recommendation and give our task formulation. 

\subsection{Collaborative Knowledge Graph}

The collaborative knowledge graph (CKG) encodes user-item interactions and item knowledge as a unified relational graph.
Let $\mathcal{U} \in \mathbb{R}^{M}$, $\mathcal{I} \in \mathbb{R}^{N}$ denote the sets of users and items, respectively.
The historical user-item interaction matrix $\mathcal{Y}=\left\{y_{u i} \mid u \in \mathcal{U}, i \in \mathcal{I}\right\}$ is defined according to users’ implicit feedback, where each entry $y_{ui}=1$ indicates there is an interaction between $u$ and $i$ and otherwise, $y_{ui}=0$.
In addition to user-item interactions, we also have an external knowledge graph (KG) that profiles item knowledge, which consists of real-world item and attribute entities and relations among them.
Inspired by~\cite{wang2019kgat}, we integrate rich item knowledge and the interaction data into a collaborative knowledge graph.
Let $\mathcal{E}$ and $\mathcal{R}$ denote the entity and relation sets in the external KG.
We first establish an item-entity alignment function $\psi: \mathcal{I} \to \mathcal{E}$ to map items in the interaction data into KG entities, in which each item $i \in \mathcal{I}$ is mapped to a corresponding item entity $e \in \mathcal{E}$ in the KG.  
Then, CKG can be formulated as $\mathcal{G}=\left\{(h, r, t) \mid h, t \in \mathcal{E}^{\prime}, r \in \mathcal{R}^{\prime}\right\}$, where $\mathcal{E}^{\prime}=\mathcal{E} \cup \mathcal{U}$, and $\mathcal{R}^{\prime}=\mathcal{R} \cup \{\mathbb{I}({y_{ui})}\}$.
$\mathbb{I}(\cdot)$ is an edge indicator that denotes the observed edge between user $u$ and item $i$ when $y_{ui}=1$, i.e., $\mathbb{I}({y_{ui})}=1$ when ${y_{ui}}=1$, otherwise $\mathbb{I}({y_{ui})}=0$.
Each triplet $(h, r, t)$ describes that there is a link $r$ from head entity $h$ to tail entity $t$, which contains rich semantic relations among diverse types of real-world entities.
For example, \textit{(Michael Jackson, SingerOf, Beat it)} states the fact that \textit{Michael Jackson} is the singer of the song \textit{Beat it}.

\subsection{Counterfactual Explanation for Recommendation }\label{sec:cf}

The counterfactual explanation is defined as a minimal perturbation set of original sample that, if applied, would result in the opposite prediction of the sample~\cite{mothilal2020explaining}.
In this work, we focus on finding the attribute-based counterfactual explanations, in which the minimal perturbation set is defined as a set of item attributes, e.g., \textit{genre}, \textit{brand}.
Specifically, suppose $f_R$ is a Top-$K$ recommendation model that gives the recommendation list $\boldsymbol{Q}_u$ with length $K$ for a user $u$, and we say $i \in \boldsymbol{Q}_u$ if item $i$ is recommended.
For a user-item pair $(u,i)$, we aim to search for a minimal item attributes set $\Delta_{ui}=\left\{a_{ui}^{1}, \cdots, a_{ui}^{r}\right\}$.
Each item attribute $a_{ui}^{i} \in \Delta_{ui}$ is an attribute entity, which contains real-world semantic that describes the item.
With the recommendation model $f_R$, and the attributes set $\Delta_{ui}$, our optimization goal is to estimate whether applying the $\Delta_{ui}$ on the recommended item $i$ results in replacing the $i$ with a different item $j$ for user $u$.
If the optimization goal meet, the attributes set $\Delta_{ui}$ is termed a \emph{counterfactual explanation} that flips the recommendation result and $j$ is the \emph{counterfactual item} for the original item $i$. 
Meanwhile, the counterfactual explanation $\Delta_{ui}$ is \emph{minimal} such that there is no smaller set $\Delta_{ui}^{\prime} \in \mathcal{E}$ satisfying $|\Delta_{ui}^{\prime}| < |\Delta_{ui}|$ when $\Delta_{ui}^{\prime}$ also meets the optimization goal.
With the optimized $\Delta_{ui}$, we can generate the counterfactual explanation for recommending item $i$ to user $u$, which takes the following form:

\vspace{0.1in}
\fbox{%
    \parbox{0.45\textwidth}{%
 Had a minimal set of attributes [$\Delta_{ui}$(s)] been different for item $i$, the recommend item would have been $j$ instead.
    }
}
\vspace{0.1in}

\subsection{Task Formulation}
Given the collaborative knowledge graph $\mathcal{G}$, we aim to construct a counterfactual explanation model that exploits rich relations among $\mathcal{G}$ to produce counterfactual items for original recommend items.
We formulate our counterfactual explanation model as:
\begin{equation}
    j\sim \pi_E(u,i, f_R,\mathcal{G}|\Theta_E),
\end{equation}
where $\pi_E(\cdot)$ is the counterfactual explanation model parameterized with $\Theta_E$, $f_R$ is the Top-$K$ recommendation model that produces the recommendation results.
The counterfactual explanation model generates empirical distribution over items among  $\mathcal{G}$ to yield counterfactual item $j$ for the recommended item $i$, which is expected to meet the counterfactual goal with a minimal set of item attributes that reverses the recommendation result. 

We also aim to use counterfactual items produced by $\pi_E$ to improve the recommendation model $f_R$.
Conceptually, a counterfactual item can offer high-quality negative signals to the recommendation, since it owns attributes that make the positive item not match the user's preference anymore.
Inspired by pairwise ranking learning~\cite{narasimhan2020pairwise}, we pair one positive user-item interaction with one counterfactual item to aggregate counterfactual signals into the recommendation model $f_R$.
Formally, 
\begin{equation}
    \min_{\Theta_{R}}{-\ln \sigma(f_R(u,i|\Theta_{R})-f_R(u,j|\Theta_{R}))}, \forall j \sim \pi_E(\Theta_E)
\end{equation}
where $f_R(\cdot)$ is the recommendation model parameterized with $\Theta_R$ and $\sigma(\cdot)$ is the sigmoid function. 
$(u,i)$ is the positive interaction that satisfies $\exists r\in \mathcal{R}^{\prime}, s.t. (u,r,i) \in \mathcal{G}$.
This formulation encourages positive items to receive higher prediction scores from the target user than counterfactual items. 
As such, the recommendation model not only produces recommendation results for counterfactual explanation model training, but is also co-trained with the counterfactual explanation model by interactively aggregating counterfactual signals to enhance the recommendation.

\section{Model Framework}

\begin{figure*}
    \centering
    \includegraphics[width=\textwidth]{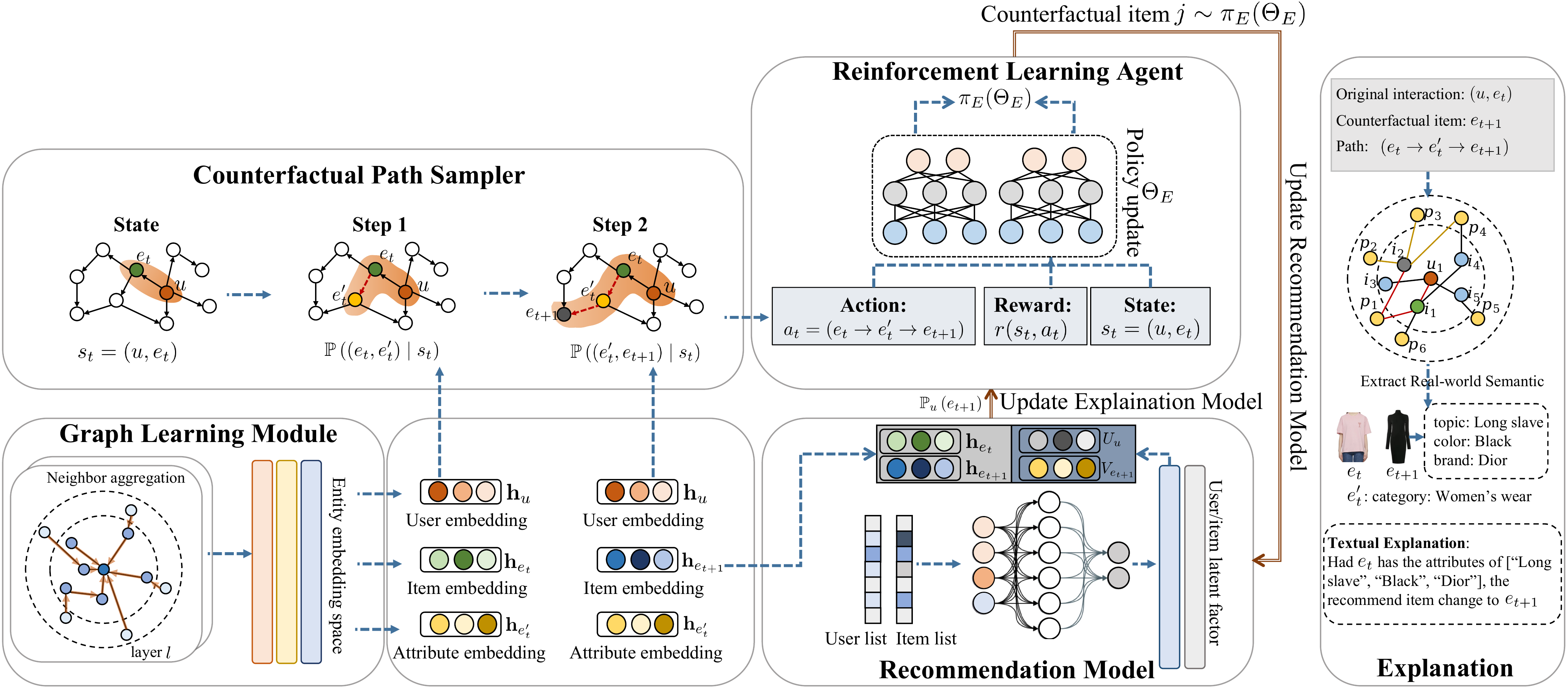}
    \caption{Framework of \textit{CERec}: 
    \textit{graph learning module} learns embeddings of users, items, and attributes entities from a CKG;  
    \textit{recommendation model} generates ranking scores of items for users;
    \textit{counterfactual path sampler} uses entity embeddings to sample paths as actions for reinforcement learning agent; 
    \textit{reinforcement learning agent} learns the explanation policy by optimizing the cumulative counterfactual rewards of deployed actions from the sampler.
    The learnt explanation policy outputs counterfactual items for original user-item interactions. 
    Meanwhile, paths connecting original items and counterfactual items from path sampler are saved to retrieve counterfactual item attributes.
    Finally, counterfactual explanations are generated by abstracting real-world semantics of counterfactual items and counterfactual item attributes.
    }
    \label{fig:framework}
\end{figure*}

We now introduce our counterfactual explainable recommendation (\textit{CERec}) framework that leverages rich relations among a collaborative knowledge graph (CKG) to generate attribute-based counterfactual explanations meanwhile providing improved recommendations.
Our \textit{CERec} consists of three modules - one recommendation model, one graph learning module and the proposed counterfactual explanation model. 
The recommendation model generates ranking scores and is co-trained with the counterfactual explanation model by interactively aggregating the produced counterfactual items.
The graph learning module embeds users, items, and attributes entities among a given CKG as embedding vectors.
The counterfactual explanation model conducts effective path sampling based on entity embeddings and ranking scores to discover high-quality counterfactual items.
Two main parts are performed in our counterfactual explanation model:
1) counterfactual path sampler: uses entity embeddings to sample paths as actions for reinforcement learning agent; 
2) reinforcement learning agent: learns explanation policy by optimizing the cumulative counterfactual rewards of deployed actions from the sampler.
We depict the framework of \textit{CERec} in Figure~\ref{fig:framework}. 
We introduce the recommendation model and the graph learning module and give an overview of our counterfactual explanation model as below.

\subsection{Recommendation Model}\label{sec:Recommendation Model}

We now present the recommendation model $f_R$ that uses user and item latent factors for Top-$K$ recommendation.
Here we employ the pairwise learning-to-rank model CliMF~\cite{10.1145/2365952.2365981} as the recommendation model.
The CliMF initializes IDs of users and items as latent factors, and updates latent factors by directly optimizing the Mean Reciprocal Rank (MRR) to predict ranking scores of items for users. 
It is worth noting that $f_R$ can be any model as long as it takes users' and items' embeddings as part of the input and produce ranking results, which makes our counterfactual explanation framework applicable to a broad scope of models, e.g., neural networks~\cite{wang2019heterogeneous}.
Specifically, we firstly map users and items into latent factors with the recommendation model,
\begin{equation}\label{eq:uv}
    f_R({u,i})=U_{u}^{\top} V_{i}
\end{equation}
where $U_{u}$ denotes $d$-dimensional latent factors for user $u$, and $V_{i}$ denotes $d$-dimensional latent factors for item $i$. 
We use the pairwise Mean Reciprocal Rank loss~\cite{10.1145/2365952.2365981} to define our objective function to optimize parameters $\Theta_R$ as below. 
\begin{equation}\label{eq:Recommendation Model}
\footnotesize
 \mathcal{L}_R=\min_{\Theta_{R}} \sum_{(u,i)\in \mathcal{Y}} [\ln \sigma\left(f_R({u,i})\right)+\sum_{j=1}\ln \left(1-\sigma \left(f_R({u,j})-f_R({u,i})\right)\right)]
\end{equation}
where $\mathcal{Y} \in \mathbb{R}^{M\times N}$ denotes the historical user-item interaction matrix, $\sigma(\cdot)$ is the sigmoid function and $j~\sim \pi_E(\Theta_E)$ is the counterfactual item generated from our counterfactual explanation model. 
By optimizing this loss function, we can get the 
ranking score for each item $i$ from user $u$ based on $U$ and $V$.
Formally, 
\begin{equation}\label{eq:probability}
\mathbb{P}_u \left(i \right)=\frac{\exp\left(U_{u}^{\top} V_{i}\right)}{\sum_{k=1}^{K} \exp\left(U_{u}^{\top} V_{k}\right)}
\end{equation}
where $K$ is the length of user $u$'s recommendation list $\boldsymbol{Q}_u$, $\mathbb{P}_u \left(i \right)$ indicates the ranking score of an item $i$ in $u$'s recommendation list (i.e., with $K$ items).

\subsection{Graph Learning Module}\label{sec:graph_emb}

The graph learning module (GLM) learns users, items and attributes representations (i.e., embeddings) from the given collaborative knowledge graph $\mathcal{G}$.
The learnt embeddings are deployed into our counterfactual explanation model to:
(1) calculate the importance scores of user, item and attribute entities to form sampling paths as actions for counterfactual path sampler;
(2) calculate the similarities among item entities to define the counterfactual rewards of deployed actions for reinforcement learning agent.

Inspired by recent advances in Graph Neural Networks (GNNs)~\cite{hu2020heterogeneous,wang2019heterogeneous} for graph data representation, we employ GraphSage~\cite{hamilton2017inductive} in our GLM to learn representations for users, items and attributes entities.
For an entity $e \in \mathcal{G}$, the Graphsage aggregates the information propagated from its neighbors $\mathcal{N}_e$ to learn $e$'s representation. 
As a user entity would connect with entities whose type is the item, i.e., $\mathcal{N}_e \in \mathcal{I}$, the learnt user embeddings capture the influence of historical user-item interactions.
Analogously, item entities connect with item attribute entities such that the learnt item embeddings absorb context information from item attributes. 
In particular, we firstly initialize entity representations at the $0$-th layer of GraphSage with Multi-OneHot~\cite{zhang2019stylistic} by mapping entity IDs into embeddings, where the embedding for an entity $e$ is denoted by $\mathbf{h}_e^{0}$.
Then, at the $l$-th graph convolutional layer, an entity $e$ receives the information propagated from its neighbors to update its representation, as:
\begin{equation}
    \mathbf{h}_e^{(l)}=\delta \left(\mathbf{W}^{(l)}\left(\mathbf{h}_e^{(l-1)} \| \mathbf{h}_{\mathcal{N}_{e}}^{(l-1)}\right)\right)
\end{equation}
where $\mathbf{h}_e^{(l)} \in \mathbb{R}^{d_l}$ is the embedding of an entity $e$ at layer $l$ and $d_l$ is the embedding size; $\mathbf{W}^{(l)}\in \mathbb{R}^{d_l \times 2d_{l-1}}$ is the weight matrix, $||$ is the concatenation operation and $\delta(\cdot)$ is a nonlinear activation function as LeakyReLU~\cite{xu2020reluplex}.
$\mathbf{h}_{\mathcal{N}_{e}}^{(l-1)}$ is the information propagated from $e$’s neighbors set $\mathcal{N}_e$, as:
\begin{equation}
    \mathbf{h}_{\mathcal{N}_{e}}^{(l-1)}=\sum_{e^{\prime} \in \mathcal{N}_{e}} \frac{1}{\sqrt{\left|\mathcal{N}_{e}\right|\left|\mathcal{N}_{e^{\prime}}\right|}} {\mathbf{h}_{e^{\prime}}}^{(l-1)}
\end{equation}
where $\mathcal{N}_{e}=\left\{e^{\prime} \mid \left(e, e^{\prime}\right) \in \mathcal{G} \right\}$ is a set of entities connected with $e$.

Having obtained the representations $\mathbf{h}_e^{(l)}$ at each graph convolutional layer $l \in \{1,\cdots, L\}$, we adopt layer-aggregation mechanism~\cite{xu2018representation} to concatenate embeddings at all layers into a single vector, as follows: 
\begin{equation}\label{eq:graph_embedding}
\mathbf{h}_e=\mathbf{h}_e^{(1)} + \cdots + \mathbf{h}_e^{(L)}
\end{equation}
where $\mathbf{h}_e^{(i)}$ is the embedding for an entity $e$ at the $i$-th layer. 
By performing layer-aggregation, we can capture higher-order propagation of entity pairs across different graph convolutional layers. 
After stacking $L$ layers, we obtain the final representation for each entity among the CKG.
Note that in the following, we use $\mathbf{h}_{u}$ to denote the embedding of user entity $u$, while $\mathbf{h}_{e_t}$ is the item embedding for item entity $e_t$ and $\mathbf{h}_{e^\prime_t}$ is the embedding for item attribute entity $e^\prime_t$.

\subsection{Counterfactual Explanation Model}

Our counterfactual explanation model contains two main parts: the counterfactual path sampler and the reinforcement learning agent.
The counterfactual path sampler performs two-step attention on users, items, and attributes embeddings from GLM to search for a high-quality path as action $a_t$ for each state $s_t$. 
Then action $a_t$ and state $s_t$ are fed into our reinforcement learning agent to learn the explanation policy.
Based on ranking scores produced by recommendation model and item embeddings, the reinforcement learning agent learns the reward $r(s_t,a_t)$ at state $s_t$ and updates the explanation policy $\pi_{E}(\Theta_E)$ accordingly.
Finally, based on the learnt explanation policy $\pi_{E}$ and path histories, our \textit{CERec} generates attribute-based counterfactual explanations for recommendations.
We detail our counterfactual explanation model in the next section.

\section{Reinforced Learning for Counterfactual Explanation Model}

In this section, we introduce our counterfactual explanation model assisted by GLM and recommendation model to generate explanation policy $\pi_E$ over reinforcement depth $T$. 
Our counterfactual explanation model contains two main parts:
\textit{counterfactual path sampler} (CPS) performs attention mechanisms on entities among CKG to sample paths as actions for reinforcement learning agent; 
\textit{reinforcement learning agent} learns the explanation policy $\pi_E$ by optimizing cumulative counterfactual rewards of the sampled actions from CPS.
We now introduce each part in our counterfactual explanation model as follows.

\subsection{Counterfactual Path Sampler}

The counterfactual path sampler (CPS) conducts path exploration over CKG to sample paths as actions for the latter explanation policy learning. 
The basic idea is to, conditioned on the target user, start from the recommended item, learn to navigate to its item attribute, and then yield the potential counterfactual item along the sampling paths.
In practice, to conduct such higher-order path sampling, large-scale CKGs are required since they encode rich relations (i.e., more than one-hop connectivity) among users, items, and item attributes.
As a result, counterfactual items are sampled from higher-hop neighbors of target user-item interactions to form the candidate action space. 
However, learning counterfactual explanations from the whole candidate action space is infeasible since the space would cover potentially enormous paths. 
Thus, our CPS is designed to reduce the candidate action space by filtering out irrelevant paths and selecting important paths for later explanation policy learning. 
Here, we employ attention mechanisms to calculate the importance of the visited entity condition on the source entity to generate sampling paths.
We now introduce our CPS $\mathbb{P}(a_t|s_t)$ that samples paths as actions $a_t$ at each state $s_t$ using attention mechanisms.

At each state $s_t$, the CPS produces a path of the given CKG as action.
Formally, $a_t \sim \mathbb{P}(a_t|s_t)=(e_t \rightarrow e_t^\prime \rightarrow e_{t+1})$ is the path that roots at an item $e_t$ towards another item proposal $e_{t+1}$, where $(e_t , e_t^\prime ), (e_t^\prime , e_{t+1})\in \mathcal{G}$ and $e_t, e_{t+1}\in I$ are connected via the item attribute $e_t^\prime$. 
The action $a_{t}=\left(e_{t} \rightarrow e_{t}^{\prime} \rightarrow e_{t+1}\right)$ is generated by the two-step path sampling: 
1) choose an outgoing edge from $e_{t}$ to the internal item attribute entity $e_{t}^{\prime}$; 
2) determine the third entity $e_{t+1}$ conditioned on $e_{t}^{\prime}$. 
We separately model the confidence of each exploration step into two attention mechanisms, i.e., $\mathbb{P}\left(\left(e_{t}, e_{t}^{\prime}\right) \mid s_{t}\right)$ and $\mathbb{P}\left(\left(e_{t}^{\prime}, e_{t+1}\right) \mid s_{t}\right)$. 

The first attention mechanism $\mathbb{P}\left(\left(e_{t}, e_{t}^{\prime}\right) \mid s_{t}\right)$ specifies the importance of item attributes for $e_{t}$, which are sensitive to the current state $s_t=(u, e_t)$, i.e., user $u$ and item $e_t$.
Formally, for each outgoing edge from $e_t$ to its item attribute $e_t^\prime \in \mathcal{N}_{e_t}$, we first obtain the embeddings of item $e_t$, attribute $e_t^\prime$ and user $u$ from Eq~\eqref{eq:graph_embedding}, which are denoted by $\mathbf{h}_{{e}_t}$, $\mathbf{h}_{e_t^\prime}$ and $\mathbf{h}_u$.
Then, the importance score of item attribute $e_{t}^{\prime}$ is calculated by:
\begin{equation}
    \alpha^{(1)}\left(e_{t}, e_{t}^{\prime}\right)=\mathbf{h}_u^{\top} \delta \left(\mathbf{h}_{{e}_t} \odot \mathbf{h}_{{e}_t^\prime}\right)
\end{equation}
where $\alpha^{(1)}$ is the attention score at the first attention mechanism. 
$\odot$ is the element-wise product and $\delta(\cdot)$ is LeakyReLU~\cite{xu2020reluplex}.
Thereafter, we employ a softmax function to normalize the scores of all neighbors of $e_t$ as:
\begin{equation}\label{eq:first_step}
\mathbb{P}\left(\left(e_{t}, e_{t}^{\prime}\right) \mid s_{t}\right)=\frac{\exp \left( \alpha^{(1)}\left(e_{t}, e_{t}^{\prime}\right)\right)}{\sum_{e_{t}^{\prime \prime} \in \mathcal{N}_{e_{t}}} \exp \left( \alpha^{(1)}\left(e_{t}, e_{t}^{\prime \prime}\right)\right)}
\end{equation}
where $\mathcal{N}_{e_{t}}$ is the neighbor item attributes set for $e_{t}$.

Having selected item attribute $e_t^\prime$, we employ another attention mechanism $\mathbb{P}\left(\left(e_{t}^{\prime}, e_{t+1}\right) \mid s_{t}\right)$ to decide yield which item from its neighbors $\mathcal{N}_{e_t^\prime}$ as the counterfactual item proposal.
We firstly calculate the attention score of $e_{t+1} \in \mathcal{N}_{e_t^\prime}$ based on attribute embedding of $e_t^\prime$, item embedding of $e_{t+1}$ and user embedding of $u$, as:
\begin{equation}
    \alpha^{(2)}\left(e_t^\prime, e_{t+1}\right)=\mathbf{h}_u^{\top} \delta\left(\mathbf{h}_{{e}_t^\prime} \odot \mathbf{h}_{{e}_{t+1}}\right)
\end{equation}
where $\alpha^{(2)}$ is the attention score at the second attention mechanism. 
$\mathbf{h}_u$, $\mathbf{h}_{e_t^\prime}$ and $\mathbf{h}_{{e}_{t+1}}$ are embeddings of user $u$, attribute $e_t^\prime$ and item ${e}_{t+1}$.
Then we normalize attention scores for all item neighbors in $\mathcal{N}_{e_t^\prime}$ to generate the selection probability of item $e_{t+1}$.
Moreover, since we care for generating valid counterfactual items as proposals, we filter out irrelevant items that do not meet the counterfactual goal, i.e., those being recommended for $u$ in the recommendation list $\boldsymbol{Q}_u$.
Formally, 
\begin{equation}\label{eq:second_step}
\small
        \mathbb{P}\left(\left(e_{t}^{\prime}, e_{t+1}\right) \mid s_{t}\right)= 
    \begin{cases}
        \frac{\exp \left(\alpha^{(2)}\left(e_{t}^{\prime}, e_{t+1}\right)\right)}{\sum_{e_{t+1}^{\prime \prime} \in \mathcal{N}_{e_{t}^{\prime}}} \exp \left(\alpha^{(2)}\left(e_{t}^{\prime}, e_{t+1}^{\prime \prime}\right)\right)}, & e_{t+1} \notin \boldsymbol{Q}_u \\
        0, & e_{t+1} \in \boldsymbol{Q}_u 
    \end{cases}
\end{equation}

Finally, the two attention mechanisms are aggregated into the CPS $\mathbb{P}\left(a_{t} \mid s_{t}\right)$ to yield the path $a_{t}=\left(e_{t} \rightarrow e_{t}^{\prime} \rightarrow e_{t+1}\right)$ as action for each state $s_t$:
\begin{equation}\label{eq:path_exploration}
    \mathbb{P}\left(a_{t} \mid s_{t}\right)=\mathbb{P}\left(\left(e_{t}, e_{t}^{\prime}\right) \mid s_{t}\right) \cdot \mathbb{P}\left(\left(e_{t}^{\prime}, e_{t+1}\right) \mid s_{t}\right)
\end{equation}
where $\mathbb{P}\left(\left(e_{t}, e_{t}^{\prime}\right) \mid s_{t}\right)$ represents the probability of stepping from $e_t$ to $e_t^\prime$ is derived from Eq.~\eqref{eq:first_step};  $\mathbb{P}\left(\left(e_{t}^{\prime}, e_{t+1}\right) \mid s_{t}\right)$ derived from Eq.~\eqref{eq:second_step} is the probability of selecting $e_{t+1}$ as the counterfactual item proposal. 
With $\mathbb{P}\left(a_{t} \mid s_{t}\right)$, 
we can generate the action $a_{t}$ for each state $s_t$ for explanation policy learning.

\subsection{Reinforcement Learning Agent}

We formulate the counterfactual explanation model as the  path-based reinforcement learning (RL) agent to discover high-quality counterfactual items.
Each action $a_t$ is a path towards candidate counterfactual item.
With the counterfactual reward $r(s_t, a_t)$ measures whether the action $a_t$ returns a valid counterfactual item for current state $s_t$.

\subsubsection{Agent}
Formally, the counterfactual explanation model is formulated as a Markov Decision
Process (MDP) $\mathcal{M}=\{\mathcal{S}, \mathcal{A}, \mathcal{P}, \mathcal{R}\}$, where
$s_t \in \mathcal{S}$ is the state absorbing the current user and the visited entity, 
$a_t \in \mathcal{A}$ is the action deposited to the current state.
$\mathcal{P}$ is the transition of states, and $\mathcal{R}$ is the reward function.
In the policy learning, the explanation policy $\pi_E(a_t|s_t)$ selects an action $a_t \in \mathcal{A}$ to take conditioning on the current state $s_t \in \mathcal{S}$, and the counterfactual explanation model receives counterfactual reward $r(s_t, a_t) \in \mathcal{R}$ for this particular state-action pair.
The final explanation policy is learnt to maximize the expected cumulative counterfactual rewards. 
We introduce these key elements for RL as follows.
\begin{itemize}
    \item $\mathcal{S}$: is a continuous state space describing a target user and the current visited item entity among the CKG.
    Formally, for a user $u$, at step $t$, the user state $s_t$ is defined as $s_t=(u, e_{t})$, where $u \in \mathcal{U}$ is a specific user and $e_t \in \mathcal{I}$ is the item entity the agent visits currently. 
    The initial state $s_0$ is $(u,i)$ and $i$ is the positive item of $u$, i.e., $y_{ui} \in \mathcal{Y}$.
    \item $\mathcal{A}$: is a discrete space containing actions available for policy learning.
    The action $a_t \in \mathcal{A}$ is a path 
    sampled from the CPS $\mathbb{P}(a_t|s_t)$.
    \item $\mathcal{P}$: is the state transition which absorbs transition probabilities of the current states to the next states.  
    Given action $a_t$ at state $s_t$, the transition to the next state $s_{t+1}$ is determined as
    $\mathbb{P}\left(s_{t+1}\mid s_{t}, a_{t}\right)\in \mathcal{P} =1$, where
    $s_{t+1}=(u,e_{t+1})$, $s_t=(u,e_t)$ and $a_t=(e_t \to e_t^\prime \to e_{t+1})$.
    \item $\mathcal{R}$: 
    $r(s_t, a_t) \in \mathcal{R}$ is the counterfactual reward measures whether the visited item $e_{t+1}$ is a valid counterfactual item by deploying action $a_t=(e_t \to e_t^\prime \to e_{t+1})$ at state $s_t$, which is defined based on the two criteria: 
    1) \textit{Rationality}~\cite{verma2020counterfactual}: 
    $e_{t+1}$ should receive a high confidence of being removed from the current user $u$'s recommendation list compared with the original item $e_{t}$; 
    2) \textit{Similarity}~\cite{DBLP:journals/corr/abs-2001-07417}: as a counterfactual explanation requires the minimal change of item attributes between counterfactual item and original item, $e_{t+1}$ should be as similar as possible with the original item $e_t$.
    The formal definition of the reward $r(s_t,a_t)$ is given by:
    \begin{equation}
    \small
    r(s_t, a_t)=\left\{\begin{array}{l}1+\cos (\mathbf{h}_{{e}_t}, \mathbf{h}_{{e}_{t+1}}), \text { if } \mathbb{P}_u \left(e_t \right) - \mathbb{P}_u \left(e_{t+1} \right)  \geq \epsilon  \\ \cos (\mathbf{h}_{{e}_t}, \mathbf{h}_{{e}_{t+1}}), \text { otherwise }\end{array}\right.
    \end{equation}
    where $\epsilon$ is the recommendation threshold determines \textit{Rationality}. $\epsilon$ is defined as the margin between ranking scores of the original item $e_{t}$ and the $K$-th item (i.e., $\boldsymbol{Q}_{u}^{K}$) in $u$'s recommendation list $\boldsymbol{Q}_{u}$, i.e., $\epsilon=\mathbb{P}_{u}({e_{t}})-\mathbb{P}_{u}({\boldsymbol{Q}_{u}^{K}})$. 
    $\cos (\mathbf{h}_{{e}_t}, \mathbf{h}_{{e}_{t+1}})$ is the cosine similarity between item embeddings $\mathbf{h}_{{e}_t}$ and $\mathbf{h}_{{e}_{t+1}}$ and is used to measure \textit{Similarity}, i.e., 
    $\cos (\mathbf{h}_{{e}_t}, \mathbf{h}_{{e}_{t+1}})=\frac{\langle \mathbf{h}_{{e}_{t}}, \mathbf{h}_{{e}_{t+1}}\rangle}{\|\mathbf{h}_{{e}_{t}}\| \| \mathbf{h}_{{e}_{t+1}}\|}$. 
    Note that $\mathbf{h}_{e_t}$ and $\mathbf{h}_{{e}_{t+1}}$ are obtained by Eq.\eqref{eq:graph_embedding}.

\end{itemize}

\subsubsection{Objective Function}

Using the trajectories $\{\mathcal{S}, \mathcal{A}, \mathcal{P}, \mathcal{R}\}$ from the agent, 
our counterfactual explanation model seeks a counterfactual explanation policy $\pi_E$ that maximizes the cumulative rewards $R(\pi_E)$ over reinforcement depth $T$:
\begin{equation}\label{eq:counterfactual explanation model}
R(\pi_E)=\mathbb{E}_{s_{0} \sim \mathcal{S}, a_{t} \sim \mathbb{P}\left(a_{t} \mid s_{t}\right), s_{t+1} \sim \mathbb{P}\left(s_{t+1} \mid s_{t}, a_{t}\right) }
{\left[\sum_{t=0}^{T} \gamma^{t} r\left(s_{t}, a_{t}\right)\right]}
\end{equation}
where $T$ is the terminal step determines reinforcement depth, $\gamma^{t}$ is a decay factor at current step $t \in T$.
$\pi_E$ is the explanation policy that produces counterfactual items for users' recommend items.
The final counterfactual explanations are formed by distilling the attributes of counterfactual items and their real-world semantics.

\section{Model Optimization}
We adopt the iteration optimization~\cite{jiang2019structured} to optimize the recommendation model and the counterfactual explanation model.
The recommendation model is initialized to provide ranking scores for counterfactual explanation model training; then the counterfactual explanation model is optimized to output high-quality counterfactual items for positive user-item interactions.
Thereafter, the recommendation model is updated by pairing one positive user-item interaction with one counterfactual item generated through the counterfactual explanation model. 
Next, we detail the recommendation model optimization and the explanation policy optimization.

\subsection{Recommendation Model Optimization}

We first initialize recommendation model by pairing one positive interaction $y_{ui} \in \mathcal{Y}$ with one unobserved item $v \in \mathcal{I}$ sampled from uniform sampling~\cite{rendle2009bpr}.
Then the recommendation model is optimized by training together with the counterfactual explanation model. 
At each iteration, the counterfactual explanation model outputs a counterfactual item $j \sim \pi_{E}(\Theta_E)$, such counterfactual item $j$ is then fed into recommendation model to update user and item latent factors $U$ and $V$ in Eq.\eqref{eq:uv}.
Finally, the recommendation model loss $\mathcal{L}_R$ in Eq.\eqref{eq:Recommendation Model} w.r.t. model parameters $\Theta_R$ is optimized by using stochastic gradient descent (SGD)~\cite{bottou2012stochastic}.

\subsection{Explanation Policy Optimization}

Since our counterfactual explanation model involves discrete sampling steps, i.e., the counterfactual path sampler, which block gradients when performing differentiation. 
Conventional policy gradient methods such as stochastic gradient descent~\cite{bottou2012stochastic} fail to calculate such hybrid policy gradients.
Thus, we solve the optimization problem through REINFORCE with baseline~\cite{sutton1999policy}.
Having obtained the policy optimization function from Eq.~\eqref{eq:counterfactual explanation model}, and the sampling function from Eq.~\eqref{eq:path_exploration},
the optimization goal is to combine the two functions and optimize them together with a REINFORCE policy gradient w.r.t. $\Theta_E$:
\begin{equation}
\begin{aligned}
\mathcal{L}_E &=\nabla_{\Theta_{E}} R(\pi_{E}) \\
& =\nabla_{\Theta_{E}} \mathbb{E}_{s_{0} \sim \mathcal{S}, a_{t} \sim \mathbb{P}(a_t|s_t), s_{t+1} \sim \mathbb{P}\left(s_{t+1} \mid s_{t}, a_{t}\right)}\left[\sum_{t=0}^{T} \gamma^{t} r(s_t, a_t)\right] \\
& \simeq \frac{1}{T} \sum_{t=0}^{T}\left[\gamma^{t} r(s_t, a_t) \nabla_{\Theta_{E}} \log \mathbb{P}\left(a_{t} \mid s_{t}\right)\right]
\end{aligned}
\end{equation}
where $\Theta_E$ are learnable parameters for our counterfactual explanation model.

\section{Experiments}

We thoroughly evaluate the proposed CERec for counterfactual explainable recommendation on three publicly available datasets.
We evaluate our CERec in terms of recommendation performance and explainability performance to particularly answer the following research questions:
\begin{itemize}
    \item Whether CERec with a counterfactual explanation model could improve the recommendation performance compared with state-of-the-art recommendation models?
    
    \item Are the counterfactual explanations generated by CERec appropriate to explain users' preferences? 

    \item How the reinforcement depth $T$ in CERec impacts the recommendation performance? 

    \item How does the training epoch impact the stability of our recommendation model?

\end{itemize}

\subsection{Experimental Setup}

\subsubsection{Dataset}
We use three publicly available datasets: \texttt{Last-FM}, \texttt{Amazon-book} and \texttt{Yelp2018}.
The statistics of these datasets are presented in Table~\ref{tab:dataset}, which depicts their recorded historical user-item interactions and the underlying knowledge graph.
The \texttt{Amazon-book}~\footnote{http://jmcauley.ucsd.edu/data/amazon} dataset is a widely used book recommendation dataset, and the \texttt{Last-FM}~\footnote{https://grouplens.org/datasets/hetrec-2011/} is a music listening dataset. 
For \texttt{Amazon-book} and \texttt{Last-FM}, we first map their recorded items into Freebase~\cite{bollacker2007freebase} entities.
Then the knowledge graphs for \texttt{Amazon-book} and \texttt{Last-FM} are built by extracting knowledge-aware facts for each item from the Freebase.
\texttt{Yelp2018}~\footnote{https://www.yelp.com/dataset} is a business recommendation dataset. 
For \texttt{Yelp2018}, we extract its item knowledge from the local business information network to construct the knowledge graph.
Each dataset is processed by the following settings: for user-item interactions, we adopt a 10-core setting, i.e., retaining users and items with at least ten interactions.
Each knowledge-aware fact (i.e., $<$user, item$>$, $<$item, attribute$>$) is represented as a conceptual edge among the collaborative knowledge graph (CKG). 
Moreover, to ensure CKG quality, we filter out infrequent entities (i.e., lower than 10 in both datasets) and retain the relations appearing in at least 50 triplets.

\begin{table}[h]
\centering
\caption{Statistics of the datasets. Density is computed by $\#Interactions/(\#Users \cdot \#Items)$.
}
\label{tab:dataset}
\resizebox{0.45\textwidth}{!}{
\begin{tabular}{l|l|c|c|c}
\toprule
\multicolumn{2}{c|}{Dataset} & \texttt{Last-FM} & \texttt{Amazon-book}  & \texttt{Yelp2018}\\
\midrule
\multirow{3}{*}{\shortstack{User-Item\\ Interaction}}
& \#Users & 23,566  & 70,679 & 45,919
\\ & \#Items & 48,123 & 24,915  & 45,538
\\& \#Interactions & 3,034,796 & 847,733  & 1,185,068
\\ & \#Density & 0.268\% & 0.048\%   & 0.057\%
\\
\midrule
\multirow{3}{*}{\shortstack{Knowledge\\ Graph}} & \#Entities & 58,266 & 88,572  & 90,961
\\ &\#Relations & 9 & 39  & 42 
\\ &\#Triplets & 464,567 & 2,557,746  & 1,853,704
\\
\bottomrule
\end{tabular}
}
\end{table}

\subsubsection{Evaluation}\label{sec:evaluation}
To evaluate the recommendation performance, 
all models are evaluated with three representative Top-$K$ recommendation metrics: Recall@$K$, Normalized Discounted Cumulative Gain (NDCG)@$K$ and Hit Ratio (HR)@$K$.
The $K$ is set as 20 by default. 
The average results w.r.t. the metrics over all users are reported
in Table~\ref{tb:recommendation} while
a Wilocoxon signed-rank test~\cite{doi:https://doi.org/10.1002/9780471462422.eoct979}
is performed in Table~\ref{tb:recommendation} to evaluate the significance. 
We evaluate the quality of explanations in terms of \textit{consistency}.
The \textit{consistency} 
measures to what extent the attributes in counterfactual explanations are appropriate to explain users' preferences.
We adopt the Precision, Recall and $F_1$ protocols follows CountER~\cite{tan2021counterfactual}, which require ground-truth attribute sets for evaluations.
As we aim to search for a minimal item attributes set that can flip the positive user-item interaction to the negative.
The ground-truth set therefore absorbs item attributes that cause the user to dislike the item.
Formally, the ground-truth set is defined as 
$\mathcal{O}_{u i} = \{o_{u i}^{1}, \cdots, o_{u i}^{p}\}$, where $o_{u i}^{p}=1$ if user $u$ has negative preference on the $p$-th attribute of item $i$; otherwise, $o_{u i}^{p}=0$.
The implementation of negative item attributes extraction is detailed in Section~\ref{sec:implement}.
On the other hand, our model produces the attributes set $\Delta_{ui}=\left\{a_{ui}^{1}, \cdots, a_{ui}^{r}\right\}$, 
which constitutes the counterfactual explanation for user-item pair $(u,i)$.
Then, for each user-item pair, the Precision, Recall and $F_1$ of $\Delta_{ui}$ with regard to  $\mathcal{O}_{u i}$ are calculated by:
\begin{equation}\label{eq:EPr}
\begin{array}{c}
    \text {Precision}=\frac{\sum_{p=1}^{r} o_{u i}^{p} \cdot I\left(a_{u i}^{p}\right)}{\sum_{p=1}^{r} I\left(a_{u i}^{p}\right)}, \text { Recall}=\frac{\sum_{p=1}^{r} o_{u i}^{p} \cdot I\left(a_{u i}^{p}\right)}{\sum_{p=1}^{r} o_{u i}^{p} } , \\
    F_1=2 \cdot \frac{\text { Precision } \cdot \text { Recall }}{\text { Precision }+\text { Recall }}
\end{array}
\end{equation}
where $I\left(a_{u i}^{p}\right)$ is an identity function such that $I(\cdot)=1$ when $a_{u i}^{p} \neq 0$ and otherwise, $I(\cdot)=0$.

\subsubsection{Baselines}
To evaluate the recommendation and explanability performance, we compare our CERec with five popular recommendation models as well as three attribute-aware baselines.
Each baseline is aligned with different evaluation tasks either for (I) recommendation evaluation or for (II) explainability evaluation.
The baseline models are listed in the following:

\begin{itemize}
    \item \textbf{NeuMF}~\cite{he2017neural} (I): 
    extends Matrix Factorization (MF) to Deep Neural Network (DNN) for modeling user-item interactions.
    
    \item \textbf{MCrec}~\cite{hu2018leveraging} (I):
    leverages DNN to model meta path-based context from Heterogeneous Information Network (HIN). 
    It propagates the context to user/item embeddings with the co-attention mechanism.
    
    \item \textbf{KGpolicy}~\cite{wang2020reinforced} (I): 
    uses a reinforcement learning agent to explore negative items over a knowledge graph (KG). 
    It trains a Bayesian Personalized Ranking model with sampled negatives for the recommendation. 
    
    \item \textbf{KGQR}~\cite{zhou2020interactive} (I): 
    models KG information with Graph Convolution Network(GCN). 
    It learns target recommendation policy with KG-enhanced state representations through the deep Q-network.
    
    \item \textbf{KGAT}~\cite{wang2019kgat} (I): 
    introduces collaborative knowledge graph (CKG) to learn node embeddings by aggregating CKG neighbors to enhance the recommendation.
        
    \item \textbf{NeuACF}~\cite{han2018aspect} (II):
    uses a DNN to learn attribute-based latent factors of users and items through meta-path based similarity matrices.
    We calculate users' preference scores on item attributes with the learnt attribute-based latent factors.
    Then we select the last-10 important attributes for each user-item pair to construct the explanation.
    
    \item \textbf{CaDSI}~\cite{wang2022causal} (II): 
    disentangles user embeddings as separated user intent chunks.
    It learns attribute-aware intent embeddings by assigning item context trained from a HIN to each user intent chunk.
    We use attribute-aware embeddings to predict users' preference on item attributes and select the last-10 important attributes as explanations.
    
    \item \textbf{RDExp} (II): 
    We randomly select 10 attributes from item attribute space for each user-item interaction and generate explanations based on the selected attributes. 
    
\end{itemize}

\subsubsection{Implementation Details}\label{sec:implement}
We train the recommendation model by the CliMF~\footnote{https://github.com/salvacarrion/orange3-recommendation} with the train/test/validate sets, which are split from user-item interactions with a proportion of 60\%/20\%/20\% of the original dataset. 
We optimize the CliMF using stochastic gradient descent (SGD)~\cite{bottou2012stochastic}. 
The same data splitting and gradient descent method are also applied in all baselines.
The three datasets for training our counterfactual explanation model are prepossessed into collaborative knowledge graphs, and the REINFORCE ~\cite{sutton1999policy} policy gradient is calculated to update the parameters. 
The REINFORCE is also applied to KGpolicy and KGQR, and same collaborative knowledge graphs are used in KGpolicy and KGAT.
The embedding size for all baselines and our CERec is fixed as $d=64$.
Two graph convolutional layers with $\{32, 64\}$ output dimensions are performed for the graph learning in our model.
All neural networks-based (i.e., DNN, GCN) baselines also keep $2$ layers.
For MCRec, NeuACR and CaDSI, we use meta-paths with the path scheme of user-item-attribute-item to ensure the model compatibility, e.g., \textit{user-book-author-book} for \texttt{Amazon-book} dataset.
For counterfactual explanation, we freeze the parameters of the trained counterfactual explanation model to produce one counterfactual item for each positive user-item interaction.
The final explanation comprises item attributes that connect with the counterfactual item in the CKG but not with the positive user-item interaction.
For explanation consistency evaluation, we construct ground-truth attributes sets using dynamic negative sampling (DNS)~\cite{zhang2013optimizing}.
We first train an attribute-aware MF model by feeding positive user-item interactions and random negative item attributes.
The DNS then uniformly draws one item attribute from the attribute space and feeds its latent factors into the MF model to predict the preference score.
Then item attributes with the highest scores are selected as negative samples to train the MF model recursively. 
Among multiple rounds of sampling, the DNS generates negative item attributes that match users' negative preferences.

The hyper-parameters of all models are chosen by the grid search, including learning rate, $L_2$ norm regularization, discount factor $\gamma$, etc.
The maximum epoch for all methods is set as $400$, while an early stopping strategy is performed (i.e., if the loss stops to increase, then terminate the model training).
We also search the reinforcement depth $T$ in $\{1, 2, 3, 4, 5\}$ (cf. Eq~\ref{eq:counterfactual explanation model}) for our model and report its effect in Section~\ref{sec:T}.

\subsection{Counterfactual Enhanced Recommendation}

In this section, we present the recommendation performance evaluation to answer the RQ1.
At each iteration, the counterfactual explanation model in our CERec produces one counterfactual item for each positive user-item interaction to recursively train the recommendation model.
Thus, we are interested to know whether incorporating counterfactual items could boost our recommendation performance.
We present the recommendation results of our CERec and baselines in Table~\ref{tb:recommendation} and discuss the main findings below.

\begin{table*}[!hbt]
 \centering
 \caption{\label{tb:recommendation}Recommendation evaluation: bold numbers are the best results, best baselines are marked with underlines.
 }
\renewcommand{\arraystretch}{1.2}
 \resizebox{0.95\textwidth}{!}{
  \begin{tabular}{|c c| c c c| c c c|c c c|}
    \hline
 &\multicolumn{10}{c|}{Top-$K$ Recommendation Evaluation} \\ [3pt]  \cline{1-11}
    \multicolumn{1}{|c|}{} & \multicolumn{1}{|c|}{Dataset} &\multicolumn{3}{c|}{\texttt{Last-FM}} &\multicolumn{3}{c|}{\texttt{Amazon-book}} &\multicolumn{3}{c|}{\texttt{Yelp2018}}\\ [3pt]  
    \cline{2-11}
   \multicolumn{1}{|c|}{Model} & \multicolumn{1}{|c|}{$K$}  & Recall@$K$ & NDCG@$K$& \multicolumn{1}{c|}{HR@$K$} & Recall@$K$  & NDCG@$K$ & \multicolumn{1}{c|}{HR@$K$} & Recall@$K$ & NDCG@$K$ & \multicolumn{1}{c|}{HR@$K$} \\  [3pt] 
   \hline
   \hline
   \multicolumn{1}{|c|}{} &20 & 0.0651 & 0.0767 & 0.2921 & 0.0799 & 0.0611 & 0.1010 & 0.0279 & 0.0380 & 0.1008\\
   \multicolumn{1}{|c|}{NeuMF} & 40 & 0.0807 & 0.0848 & 0.3409 & 0.1377 & 0.0769 & 0.2660 & 0.0357 & 0.0411 & 0.1241\\
   \multicolumn{1}{|c|}{} &60 & 0.0913 & \underline{0.0955} & 0.3788 & 0.1581 & 0.0888 & 0.3007 & 0.0389 & 0.0448 & 0.1552\\
   \multicolumn{1}{|c|}{} &80 & 0.1005 & \underline{0.1009} & 0.4267 & 0.1888 & 0.0974 & 0.3332 & 0.0411 & 0.0477 & 0.2008\\
   \hline
    \multicolumn{1}{|c|}{} &20 & 0.0770 & 0.0821 & 0.2811 & 0.0879 & 0.0666 & 0.1420 & 0.0327 & 0.0412 & 0.1234 \\
   \multicolumn{1}{|c|}{MCRec} &40 & 0.0825 & 0.0845 & 0.3124 & 0.1041 & 0.0712 & 0.2177 &0.0338	& 0.0424 & 0.1406\\
   \multicolumn{1}{|c|}{} &60 & 0.0919 & 0.0919 & 0.3888 & 0.1801 & 0.0844 & 0.2805 & 0.0429 & 0.0436 & 0.1721 \\
   \multicolumn{1}{|c|}{} &80 & 0.1112 & 0.0945 & 0.4282 & 0.2257 & 0.0957 & 0.3672 & 0.0457 & 0.0437 & 0.2205\\
   \hline
         \multicolumn{1}{|c|}{} &20 &0.0761  & 0.0689 & 0.3197 &\underline{0.1242}  & 0.0684  & \underline{0.2211}  & \underline{0.0557}  &0.0435 &0.1528 \\
   \multicolumn{1}{|c|}{KGpolicy} &40 & \underline{0.1018} & 0.0763 & 0.4091 &\underline{0.1716}  & 0.0805  & \underline{0.2947}  &\underline{0.0649}  &0.0460  &0.2588  \\
   \multicolumn{1}{|c|}{} &60 & \underline{0.1179} & 0.0815 & 0.4639 & \underline{0.2048}  &0.0879  & \underline{0.3417}  &\underline{0.0832}  &\underline{0.0547}  &\underline{0.3300} \\
   \multicolumn{1}{|c|}{} &80 & \underline{0.1302} & 0.0855 & 0.5006 &0.2315  &0.0935  & 0.3762  &\underline{0.0883} &\underline{0.0618}  &\underline{0.3841} \\
   \hline
      \multicolumn{1}{|c|}{} &20 &0.0821  &0.0856 &0.3162 & 0.1076 & 0.0787 & 0.2182 & 0.0417 & 0.0458 & 0.1621\\
   \multicolumn{1}{|c|}{KGQR} &40 & 0.0932 & \underline{0.0923} & \underline{0.4229} & 0.1652 & \underline{0.0891} & 0.2358 & 0.0456 & 0.0499 & 0.1899 \\
   \multicolumn{1}{|c|}{} &60 &0.0999 & 0.0939 & \underline{0.5228} & 0.1987 & 0.0912 & 0.3077 & 0.0512 & 0.0535 & 0.2172\\
   \multicolumn{1}{|c|}{} &80 &0.1132 & 0.0947 & 0.5323 & 0.2212 & 0.0942 & 0.3531 & 0.0557 & 0.0587 & 0.2566\\
   \hline
         \multicolumn{1}{|c|}{} &20 &\underline{0.0870}  & \underline{0.0897} & \underline{0.3292}& 0.0801  & \underline{0.0791} &0.2006 & 0.0444 & \underline{0.0472} & \underline{0.2341}\\
   \multicolumn{1}{|c|}{KGAT} &40 &0.0962  & 0.0918 & 0.3762 & 0.1435 & 0.0812 &0.2634 & 0.0469  & \underline{0.0511} & \underline{0.2666}\\
       \multicolumn{1}{|c|}{} &60 &0.1083  & 0.0937 & 0.4515 & 0.1766  & \underline{0.0922} & 0.3244 & 0.0501  & 0.0546 & 0.3015\\
   \multicolumn{1}{|c|}{} &80 & 0.1232 & 0.0939 & \underline{0.5464} & \underline{0.2378}  & \underline{0.0985} & \underline{0.3921} &0.0544  & 0.0577 & 0.3655\\
   \hline
         \multicolumn{1}{|c|}{} &20 & \textbf{0.1015} & \textbf{0.0900} & \textbf{0.3867} &\textbf{0.1406} &\textbf{0.0803} &\textbf{0.2451} &\textbf{0.0650} & \textbf{0.0495} & \textbf{0.2515}\\
   \multicolumn{1}{|c|}{CERec} &40 & \textbf{0.1304}  & \textbf{0.0993} & \textbf{0.4801} &\textbf{0.1881} &\textbf{ 0.0926} &\textbf{0.3174} & \textbf{0.1025} & \textbf{0.0555} & \textbf{0.3549}\\
   \multicolumn{1}{|c|}{} &60 & \textbf{0.1478}  & \textbf{0.1052} & \textbf{0.5295} &\textbf{0.2203} & \textbf{0.0999} & \textbf{0.3624}& \textbf{0.1317} & \textbf{0.0639} & \textbf{0.4218}\\
   \multicolumn{1}{|c|}{} &80 & \textbf{0.1603}  & \textbf{0.1094} & \textbf{0.5628} &\textbf{0.2462} & \textbf{0.1054} & \textbf{0.3948} & \textbf{0.1572} & \textbf{0.0706} & \textbf{0.4739}\\
   \hline
   \hline
            \multicolumn{1}{|c|}{} &20 & + 16.67 & + 0.33 &+ 17.47 &+ 13.20 &+1.52 &+ 10.85 &+ 16.70 &+ 4.87 &+ 7.43\\
   \multicolumn{1}{|c|}{Improv.\%} &40 &+ 28.09 & + 7.58 &+ 13.53 &+ 9.62 &+ 3.93 &+ 6.55 &+ 57.94 &+ 8.61 &+ 33.12 \\
   \multicolumn{1}{|c|}{} &60 & + 25.36 &+ 10.16 &+ 1.28 &+ 7.57 &+ 8.35 &+ 6.06 &+ 58.29 &+ 16.82 &+ 27.82   \\
   \multicolumn{1}{|c|}{} &80 &+ 23.11 &+ 8.42 &+ 3.00 &+ 3.53 &+ 7.01 &+ 0.67 &+ 78.03 &+ 14.24 &+ 23.38 \\
  \hline 
 \end{tabular}}
\end{table*}

Our proposed CERec equipped with a counterfactual explanation model consistently outperforms all baselines across three datasets on both evaluation metrics.
For example, CERec obtains 16.67\%, 0.33\% and 17.47\% improvements for Recall@20, NDCG@20 and HR@20 respectively over the best baseline on \texttt{Last-FM} dataset.
By designing a counterfactual explanation model, CERec is capable of exploring the high-order connectivity among the CKG to generate counterfactual items for recommendation model training. 
This verifies the effectiveness of our counterfactual explanation model in producing high-quality counterfactual items that can boost the recommendation performance. 
Counterfactual items provide high-quality negative signals of user preference.
By pairing one counterfactual item with one positive user-item interaction for recommendation model training, our recommendation model learns to distinguish between positive items and negative items to generate more precise recommendations.

Among the knowledge-aware models, our CERec consistently outperforms MCRec, KGpolicy, KGQR, and KGAT.
This is mostly because these knowledge-aware baselines might not fully utilize the item knowledge by lacking the counterfactual reasoning ability as in our CERec.
All knowledge-aware baselines employ various attention mechanisms to capture users' preferences on item knowledge w.r.t. different item attributes.
However, they fail to consider the underlying mechanism that really triggered users' interactions. 
On the contrary, our CERec learns to discover the item attributes that cause the change of users' interactions, thus resulting in a promoted recommendation performance.

Our CERec achieves significant improvements over baselines on the \texttt{Yelp2018} dataset, e.g., 78.03\%, 14.24\%, and 23.38\% improvements for Recall@80, NDCG@80, and HR@80, respectively.
The \texttt{Yelp2018} dataset records a large number of inactive users that have few interactions with items.
It is well acknowledged that inferring user preference for inactive users is challenging and limit the performance of recommendation systems~\cite{10.1145/3485447.3512072}.
However, our CERec can still achieve favorable recommendation performance on the \texttt{Yelp2018}.
We attribute the improvements of our CERec to the augmenting of counterfactual items that could infer negative user preferences for inactive users and thus boost the recommendation performance. 

\subsection{Counterfactual Explanation Quality}
To answer RQ2, we evaluate the explainability of our framework by reporting the quantitative results of explanation evaluation metrics.
Then we discuss the interpretability of our generated counterfactual explanations by distilling their real-world semantics.

\subsubsection{Quantitative Results}
We report the Precision, Recall and $F_1$ (cf.~\eqref{eq:EPr}) of explanations generated by our CERec and baseline models in Table~\ref{tab:consistency} to test the \textit{consistency}.
The three evaluation metrics reflect to what extent the item attributes in explanations are appropriate to explain users’ preferences.
By analyzing Table~\ref{tab:consistency}, we have several observations:

\begin{table*}[t]
\centering
\caption{Explainability evaluation w.r.t the consistency. All numbers in the table are percentage numbers with ‘\%’ omitted. The number after $\pm$ is the standard error.}
\label{tab:consistency}
\renewcommand{\arraystretch}{1.2}
\resizebox{0.95\textwidth}{!}{
\begin{tabular}{|c| c|c|c|c|c|c|}
\hline
\multicolumn{7}{|c|}{Explanation Consistency} \\ \hline 
\multirow{2}{*}{\diagbox{Model}{Dataset}} & \multicolumn{3}{c|}{\texttt{Amazon-book}} & \multicolumn{3}{c|}{\texttt{Yelp2018}} \\ \cline{2-7}
& Precision\%      & Recall\%     & $F_1$\%     & Precision\%    & Recall\%  & $F_1$\%  \\ \hline
RDExp & $0.0196 \pm 0.00$ & $0.0090 \pm 0.00$ & $0.0117\pm 0.00$ & $0.4595\pm 0.01$ & $0.0987 \pm 0.01$  & $0.1320 \pm 0.03$  \\ \hline
NeuACF & $4.7336 \pm 1.21$ & $0.6171 \pm 0.06$ & $1.0226 \pm 0.08$ & $8.5603 \pm 1.51$ & $8.2619 \pm 1.47$ & $7.0589 \pm 1.38$\\ \hline
CaDSI & $\underline{13.2632} \pm 1.98$ & $\underline{1.0897} \pm 0.09$ &$\underline{2.8721} \pm 0.06$ & $\underline{17.7322} \pm 2.07$ & $\underline{19.0801} \pm 2.38$ & $\underline{18.3096} \pm 2.60$ \\ \hline
CERec & $\textbf{23.8969}\pm 2.84$  & $\textbf{3.1020} \pm 0.92$  & $\textbf{5.1411} \pm 0.83$ & $\textbf{45.5013} \pm 3.11$ & $\textbf{41.5931} \pm 3.74$ & $\textbf{37.2509} \pm 3.09$ \\ \hline 
\end{tabular}
}
\end{table*}

Firstly, the RDExp method 
performs very poorly on both Precision, Recall, and $F_1$.
The RDExp generates explanations by randomly sampling item attributes from the knowledge graph provided.
The poor performance of RDExp shows that randomly choosing item attributes as explanations can barely reveal the reasons for recommendations. 
This demonstrates that high-quality explanations require the appropriate choice of item attributes.
Secondly, the counterfactual explanations generated by our CERec consistently show superiority against all baselines in finding item attributes that are consistent with users' preferences.
This indicates that our CERec achieves superior explainability and can generate more relevant attribute-based explanations that truly match user preference. 
Unlike NeuACF and CaDSI that generate explanations by selecting a fixed number of item attributes indicated by their importance scores, our CERec searches a minimal set of item attributes that would flip the recommendation result.
We hence conclude that inferring item attributes with minimal changes is more appropriate to generate explanations, since they reflect simple but essential attributes that directly cause user preference changes. 
This further sheds light on the importance of conducting counterfactual explanations for recommendations.

\subsubsection{Interpretability of Counterfactual Explanations}

\begin{figure}[htbp]
\centering
\includegraphics[width=0.4\textwidth]{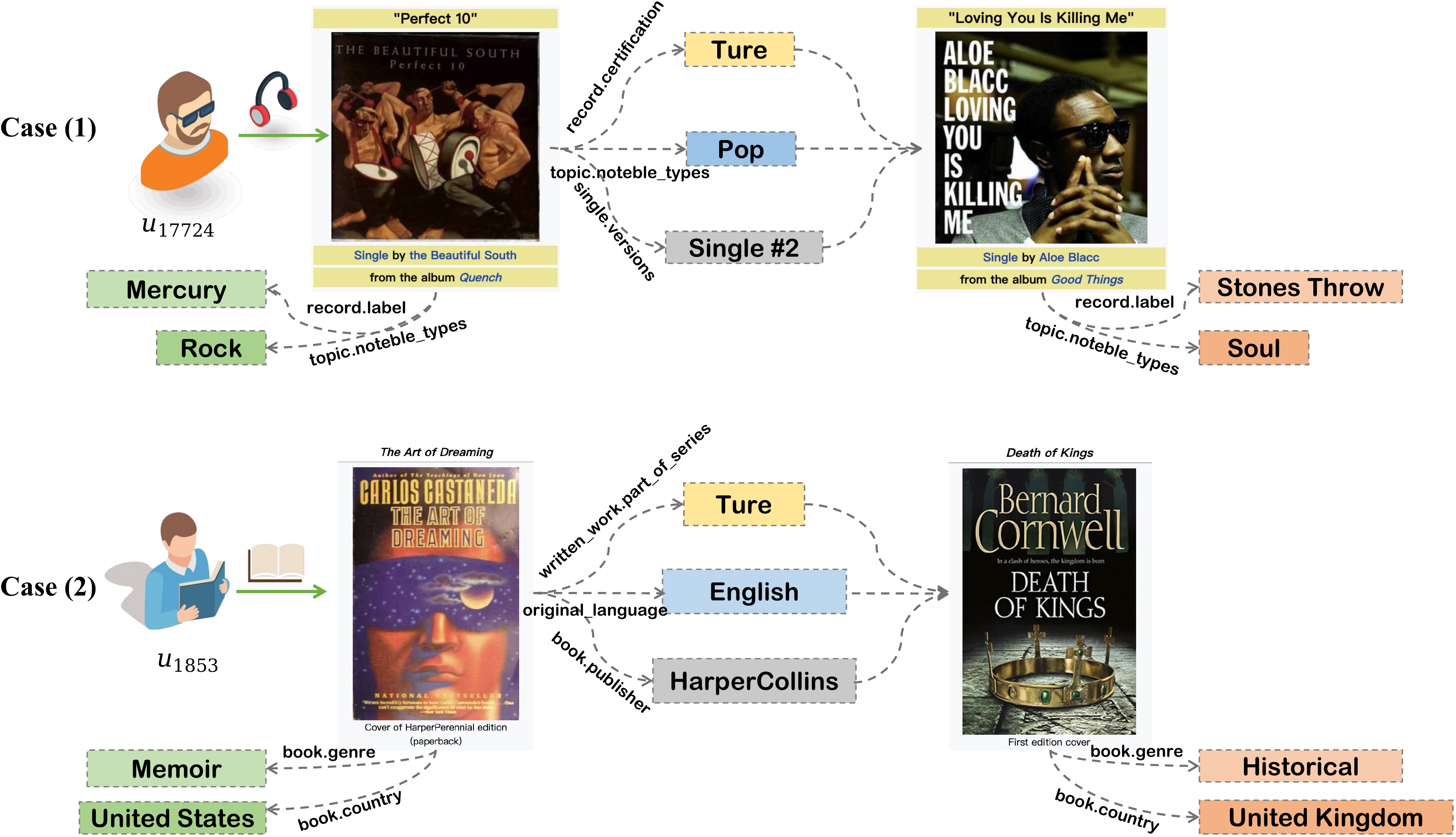}
\caption{Real cases of counterfactual reasoning paths.
}
\label{fig:path_case}
\end{figure}

We present a case study on the interpretability of counterfactual paths that root at a positive user-item interaction and end at its counterfactual item.
Each freebase entities among paths are mapped into real-world entities.
We show the real-world cases in Figure~\ref{fig:path_case} and give our observations in the following.

The first example (Case 1) comes from the \texttt{Last-FM} dataset, where user $u_{17724}$ positively interacted with ``Perfect 10'' in his listening record.
Our CERec picks the ``Loving you is killing me'' as the counterfactual item, which shares three overlapping item attributes with the ``Perfect 10''.
Conventional explainable recommendation models would pick these three common item attributes to generate explanations as ``you like Perfect 10 is because you like pop music that was certificated for its single\#2 version''. 
However, this explanation cannot answer why the ``Loving you is killing me'' with the same item attributes is actually being removed from the user's recommendation list.
By contrast, our CERec captures the counterfactual item attributes (i.e., \textit{Stones Throw} and \textit{Soul}) and generates the counterfactual explanation as ``if Perfect 10 has the label as \textit{Stones Throw} and the type as \textit{Soul}, then it will not be recommended anymore.''
This counterfactual explanation provides us with deeper insights of users' real preferences. 
For example, analyzing the attribute differences, we find that although ``Perfect 10'' shares the same type \textit{Pop} as ``Loving you is killing me'', it is actually the \textit{Rock} music published by \textit{Mercury}. 
On the contrary, the ``Loving you is killing me'' that holds the opposite music type \textit{Soul} and different polisher would potentially be disliked by the same user.

Analogously, in Case 2 from the \texttt{Amazon-book} dataset,  our CERec picks the ``Death of Kings'' as the counterfactual item for the positive interaction between $u_{1853}$ and ``The Art of Dreaming'', and uses item attributes \textit{Historical} and \textit{United Kingdom} as the counterfactual explanation to reflect $u_{1853}$'s preferences on book genre and country. 
These counterfactual explanations are more rational than conventional explanations, since they discard tedious associations of item attributes and expand explanations to counterfactual attributes.

\subsection{Parameter Analysis}
To answer RQ3, we first study how the reinforcement depth $T$ in Eq.~\eqref{eq:counterfactual explanation model} affects the recommendation performance.
We then investigate how the training epoch impacts the stability of our recommendation model.

\subsubsection{Impact of Reinforcement Depth}\label{sec:T}

The reinforcement depth $T$ determines the search space, with $T=1$ denoting that at most $1$-hop neighbors of starting entities are visited as proposals for the policy learning.
We search the number of $T$ in $\{1, 2, 3, 4, 5\}$ and report the recommendation performance of our model in Figure~\ref{fig:ablation_T}.
We have the following observations:

\begin{figure}[htbp]
\centering
\begin{minipage}[t]{0.23\textwidth}
\centering
\includegraphics[width=\textwidth]{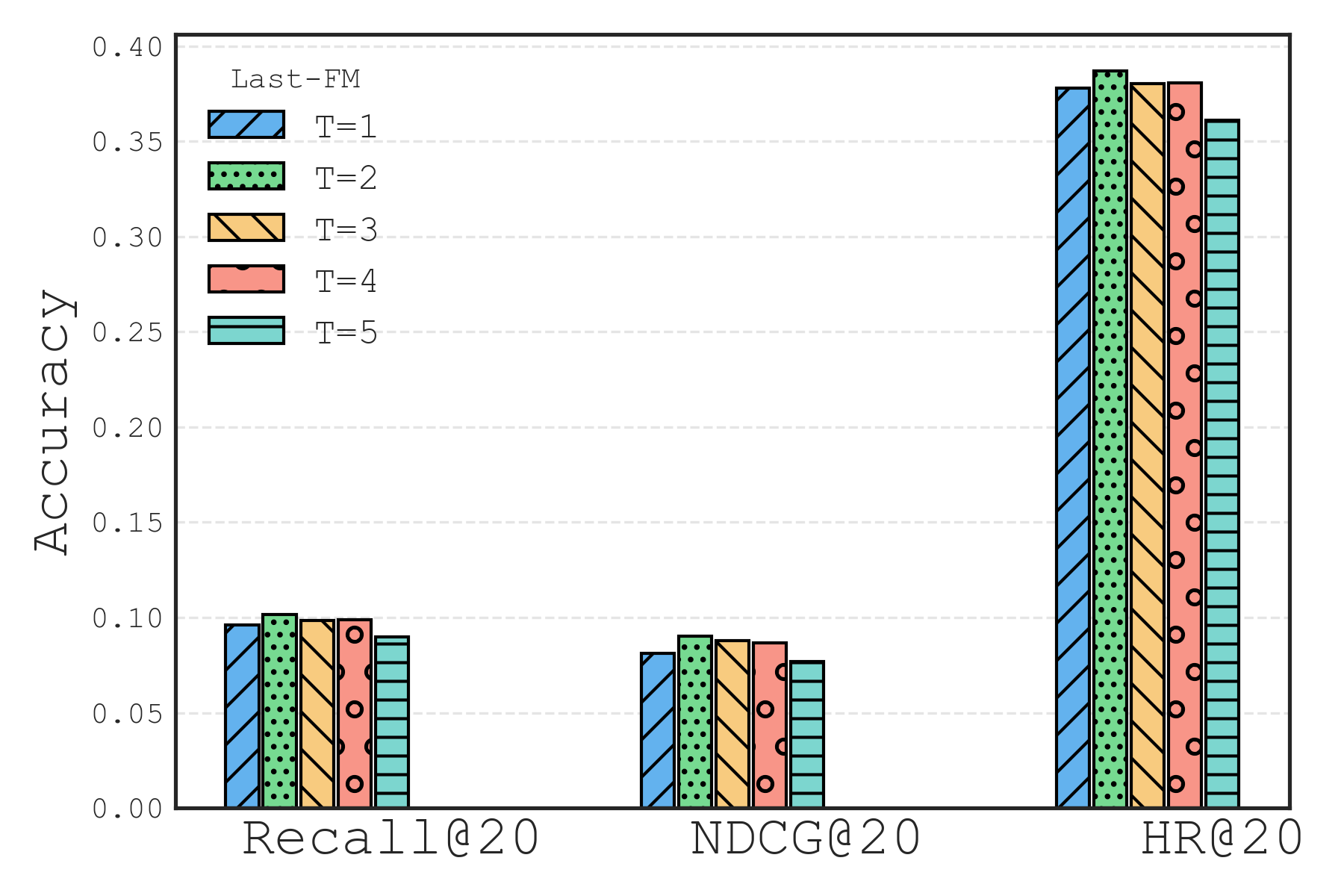}
\subcaption{\centering The impact of $T$ on \texttt{Last-FM}.}
\end{minipage}
\begin{minipage}[t]{0.23\textwidth}
\centering
\includegraphics[width=\textwidth]{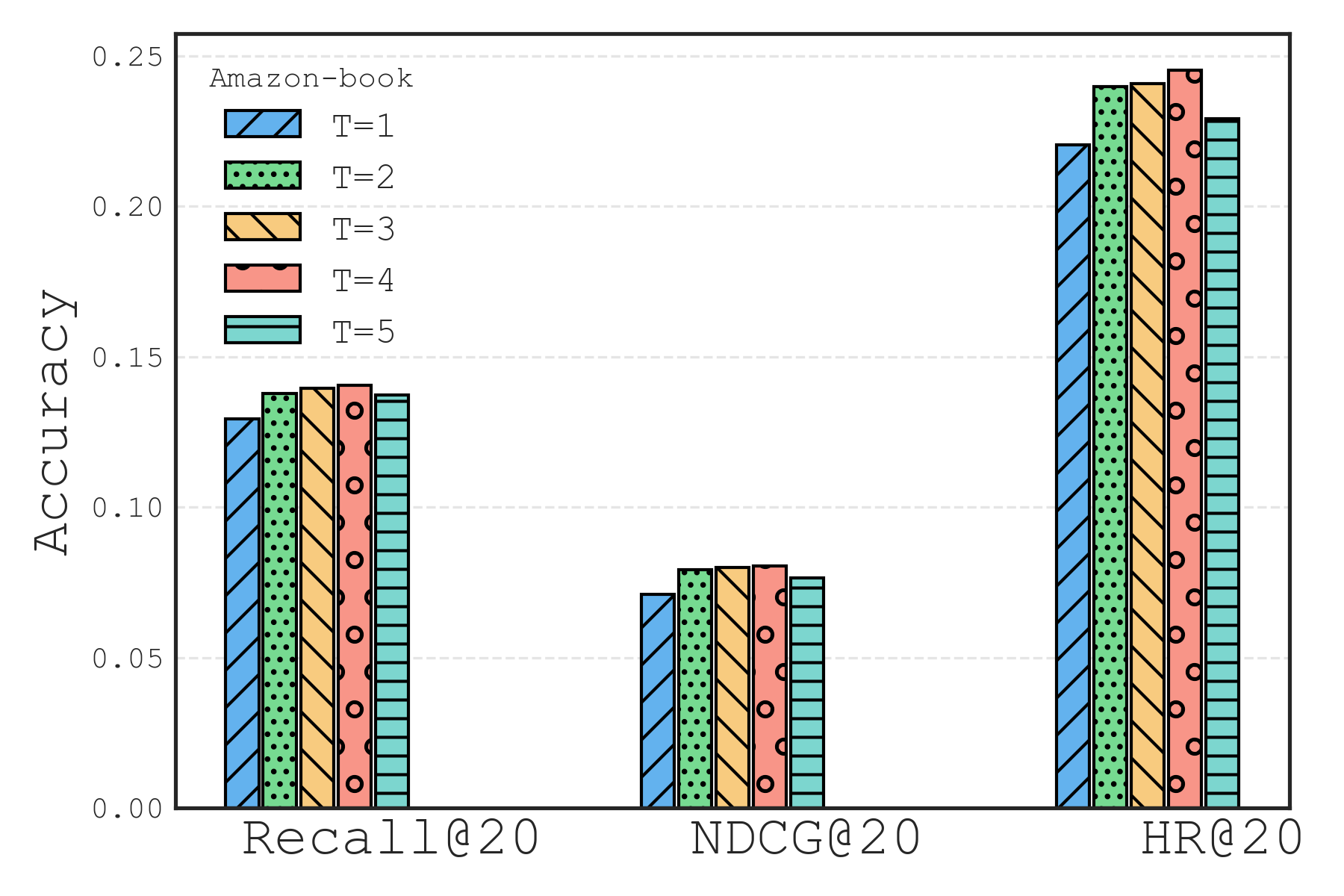}
\subcaption{\centering The impact of $T$ on \texttt{Amazon-book}.}
\end{minipage}
\begin{minipage}[t]{0.23\textwidth}
\centering
\includegraphics[width=\textwidth]{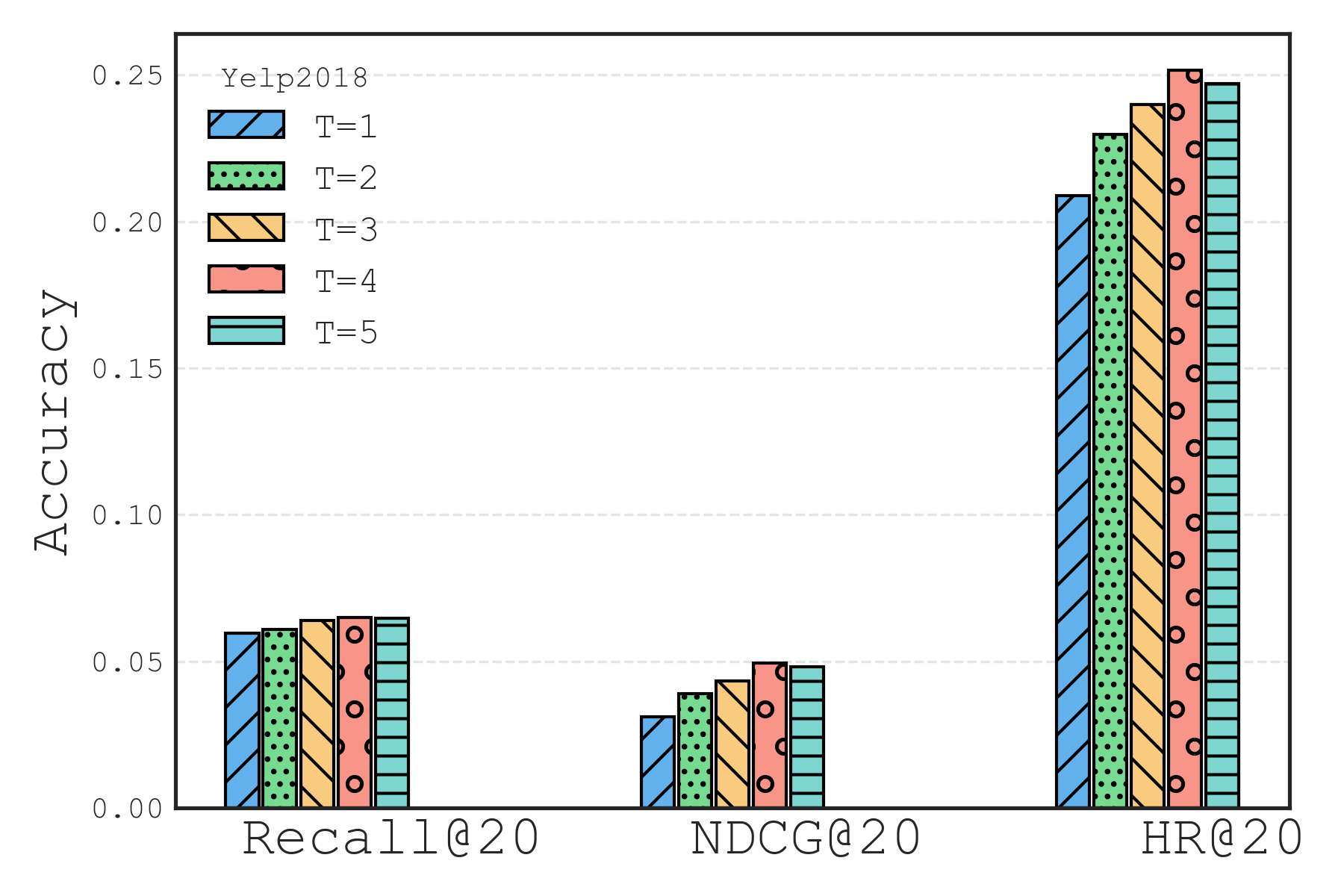}
\subcaption{\centering The impact of $T$ on \texttt{Yelp2018}.}
\end{minipage}
\caption{Impact of reinforcement depth $T$ on three datasets.}
\label{fig:ablation_T}
\end{figure}

Firstly, 
our CERec with $T=4$ yields the best performance on the \texttt{Amazon-book} and \texttt{Yelp2018} dataset, while $T=2$ gives the best results on the \texttt{Last-FM}.
The \texttt{Amazon-book} and \texttt{Yelp2018} are presented with 0.048\% and 0.057\% density and are much sparser compared with the \texttt{Last-FM} with 0.26\% density.
That means more inactive users with few item interactions are recorded in the \texttt{Amazon-book} and \texttt{Yelp2018}.
To achieve the optimal recommendation performance, the \texttt{Amazon-book} and \texttt{Yelp2018} require larger reinforcement depth (i.e., $T=4$) compared with the \texttt{Last-FM} (i.e., $T=2$).
This is because diverse counterfactual items are retrieved by increasing the reinforcement depth to help the recommendation model achieve optimal performance on the two datasets. 
This finding indicates that augmenting diverse counterfactual items could provide precise recommendations for inactive users to enhance the recommendation.
Counterfactual items offer high-quality negative signals for user preferences, and thus could help filter out negative items when providing recommendations for inactive users.
Secondly, increasing the reinforcement depth enhances the recommendation performance before reaching the peaks on both datasets.
We attribute such consistent improvements to the improved diversity of counterfactual items.
This is because higher-hop item neighbors naturally cover more items beyond those unexposed but are actually the counterfactual ones than lower-hop neighbors.
Thirdly, after peaks, increasing reinforcement depth leads to downgraded performance. 
This is because performing too many counterfactual item explorations introduces less-relevant items that may bias the recommendation results.

\subsubsection{Reward Gradient}

\begin{figure}[htbp]
\centering
\includegraphics[width=0.45\textwidth]{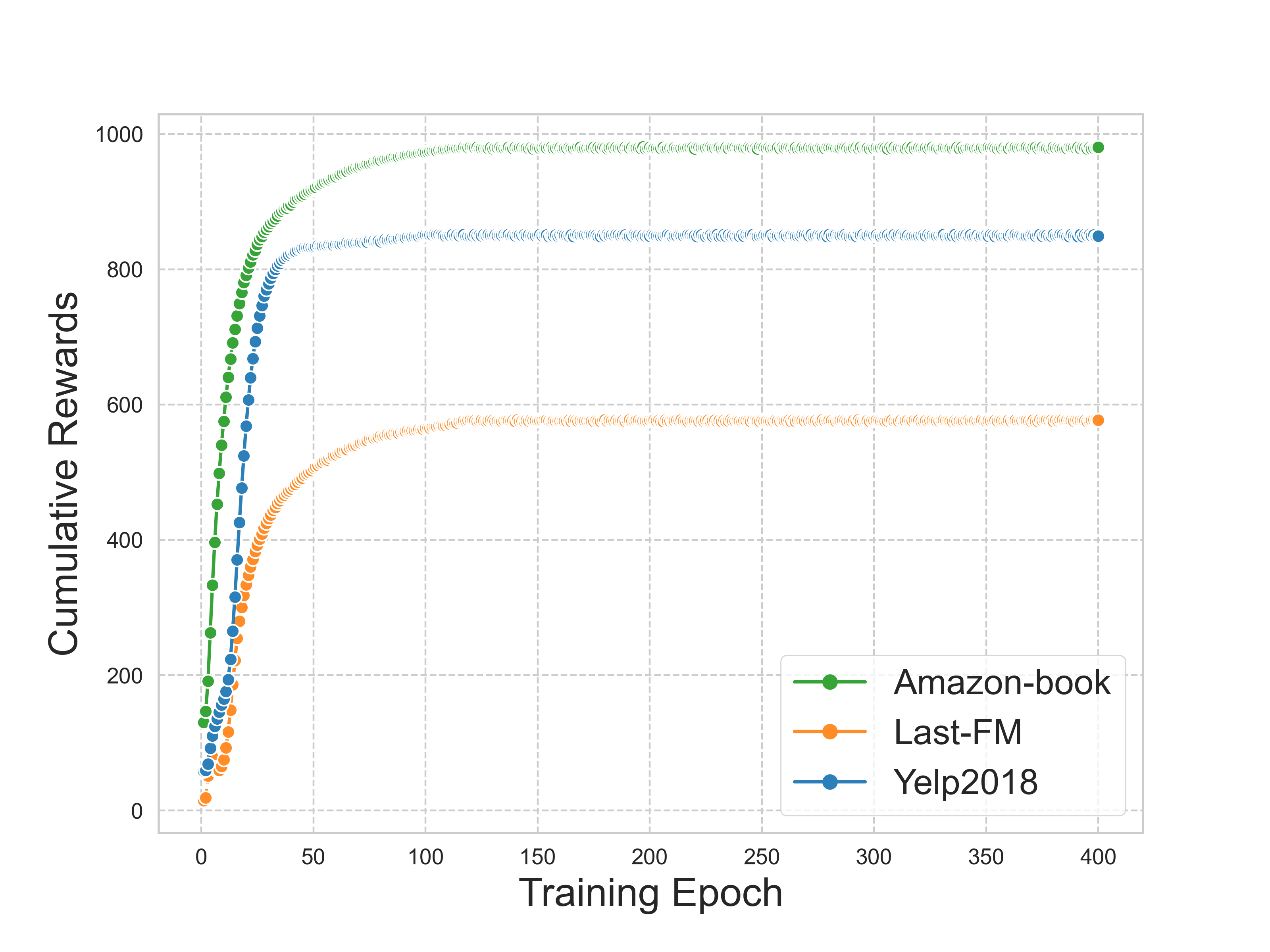}
\caption{Learning curves of cumulative rewards w.r.t. training epoch.}
\label{fig:reward}
\end{figure}

To evaluate the stability and the robustness of our model, we plot the cumulative rewards while training our model on the three datasets in Figure~\ref{fig:reward}.
Here are our observations.

First, our counterfactual explanation model gains stable cumulative rewards on both datasets along with training, i.e., the cumulative rewards increase as the epoch increases and finally reach stable states without suffering any drastic fluctuations. 
This indicates that our counterfactual explanation model enjoys stable training without losing reward gradients (e.g., gradients varnishing).
This further verifies the rationality and robustness of our proposed model.
Second, the rates of reward convergences are different across different datasets. 
For example, the cumulative rewards on \texttt{Amazon-book} and \texttt{Yelp2018} start to increase drastically at the beginning and converge at around epoch 40, while the counterpart on \texttt{Last-FM} first converges slowly and then becomes stable at around epoch 120.
This indicates that our model on \texttt{Amazon-book} and \texttt{Yelp2018} can quickly reach stable states using a small number of iterations; even \texttt{Amazon-book} and \texttt{Yelp2018} are much more sparse than \texttt{Last-FM}.
This is because we incorporate counterfactual items while training our model, which contain negative user preference signals to assist the better decision-making for inactive users. 
Besides, we explore higher-order connectivity among the CKG as side information for our model training, thus enhancing model robustness facing sparse datasets.

\section{Conclusion}

In this work, we propose CERec, a reinforcement learning-based counterfactual explainable recommendation framework over a CKG.
Our CERec is capable of generating attribute-based counterfactual explanations meanwhile provide precise recommendations.
In particular, we design a counterfactual explanation model as a reinforcement learning agent to discover high-quality counterfactual items.
The counterfactual explanation model takes paths sampled from our counterfactual path sampler as actions to optimize an explanation policy. 
By maximizing the counterfactual rewards of the deployed actions, the explanation policy is learnt to generate high-quality counterfactual items.  
In addition, we reduce the vast action space by utilizing attention mechanisms in our path sampler to yield effective paths from the CKG.
Finally, the learnt explanation policy generates attribute-based counterfactual explanations for recommendations.
We also deploy the explanation policy to a recommendation model to enhance the recommendation.
Extensive explainability and recommendation evaluations on three large-scale datasets demonstrate CERec's abilities to improve the recommendation and provide counterfactual explanations consistent with user preferences. 

\section*{Acknowledgment}

This work is supported by the Australian Research Council (ARC) under Grant No. DP200101374, LP170100891, DP220103717 and LE220100078.

\bibliographystyle{IEEEtran}
\bibliography{bibs}

\end{document}